\documentclass[11pt]{book}

\usepackage{amsmath,epsfig,graphicx,amssymb}

\usepackage{bm,units}

\newcommand{\ket}[1]{\ensuremath{\left\vert#1\right\rangle}}

\newcommand{\braket}[3]{\ensuremath{\left\langle#1\left\vert#2\right\vert#3\right\rangle}}
\newcommand{\CL}{\ensuremath{\mathcal{L}}}

\begin{document}
 \setcounter{chapter}{0}
\chapter{Testing Lorentz Invariance Violation in Quantum Gravity Theories}
\bibliographystyle{aipprocl}

 \setcounter{page}{1}

\begin{center}
{\Large {\bf H\'ector Vucetich$^\dag$
             }}  \\[3mm]
{\large {\it $^\dag$ Observatorio Astron\'omico, Universidad Nacional de
             La Plata,\\
 Paseo del Bosque S/N, (!900) La Plata, Argentina \\[3mm]
{\tt e-mail: vucetich@fcaglp.unlp.edu.ar}}}       \\[10mm]

\begin{minipage}{5in}
\centerline{{\sc Abstract}}
\medskip


Much research has been done in the latter years on the subject of
Lorentz violation induced by Quantum Gravity effects. On the
theoretical side it has been shown that both Loop Quantum Gravity and
String Theory predict that Lorentz violation can be induced at an
energy near to the Planck scale. On the other hand, most of the
experimental results in the latter years, have confirmed that the laws
of physics are  Lorentz invariant at low energy with very high
accuracy.

The inclusion of one- and two-loop contributions from a Lorentz
violating Lagrangian dramatically change the above picture: the loop
momenta run into the Planck scale and above and from the "divergent"
terms finite Lorentz violating contributions of order one arise. These
can be suppressed through suitable counterterms in the Lagrangian,
originating a strong fine tuning problem.

A brief discussion of these issues and their possible influence in
future research follows.

\end{minipage}
\end{center}

\noindent {\bf Keywords:} Please, No more, Then ten, Words

\section{Introduction}
\label{sec:Intro}

For a long time, the search of experimental clues on the nature of
quantum gravity, was dismissed as impractical, based on the simplistic
argument that those effect should appear only at energies of the order
of Planck scale ($E_P = \unit[1.2\times10^{19}]{GeV}$), far beyond
present day experimental possibilities. But recently there was a
revolutionary change in this attitude, originated in references
\cite{Kostelecky:1988zi,%
Kostelecky:1989jp,Kostelecky:1989jw,Kostelecky:1991ak,Kostelecky:1995qk}.
Here, a spontaneous breakdown of Lorentz symmetry at the Planck scale
motivated in string theory was proposed, breaking CPT symmetry and
consequently of Lorentz invariance. Later, an extension of the
standard model was developed by the same group, including all possible
CPT and Lorentz violating interactions
\cite{Colladay:1996iz,Colladay:1998fq}, which has been used to develop
sensitive tests of Lorentz symmetry.

A different approach \cite{Amelino-Camelia:1997gz,%
Amelino-Camelia:2001dy,Ahluwalia:1999aj,Amelino-Camelia:2004hm}. (See
also \cite{sorkin91}) was based on the possibility that Quantum
Gravity effects would modify the dispersion relations for particle
propagation, such as photons. These modifications in turn would change
the propagation velocity of photons, introducing delays for
particles of different energies which could be detected if these
particles would travel cosmological distances. 

Such modifications of the dispersion relations have been found in the
two most popular approaches to Quantum Gravity: Loop Quantum Gravity
\cite{Gambini:1998it,Alfaro:1999wd,Alfaro:2001rb,Alfaro:2002xz,Urrutia:2004pu}
and String theory
\cite{Ellis:1999uh,Ellis:1999sf,Ellis:1999sd,Ellis:2000dy}. These
theories predict corrections to the dispersion relations which depend
on energy in the form $(E\ell_P)^n$, with $\ell_P =
\unit[1.6\times10^{-33}]{cm}$ the Planck length scale. 

String theory has suggested another form of Lorentz violation: non
commutative field theory \cite{Seiberg:1999vs,Douglas:2001ba}, where
it is assumed that the coordinate themselves are functions from the
differential variety to a noncommutative algebra. This was originally
found in \cite{Seiberg:1999vs} where it was shown that the low energy
behavior of a bosonic string propagating in a Neveu-Schwartz
condensate can be represented by a noncommutative field theory.%
\footnote{A didactic introduction to noncommutative field theory can
be found in \cite{grandi02:PhDTh}.} 
Noncommutativity induces of course Lorentz violation and and it
constitutes one of the most interesting sources of it. 

More recently, a new approach to the testing of Quantum Gravity
effects was introduced \cite{Sudarsky:2002ue,Sudarsky:2002zy}, based
on the fact that if the dispersion relations of photons in the vacuum
are modified in the form $v(E) \neq c$, this implies a breakdown of
Lorentz invariance since such a statement can be valid at most in a
single reference frame. In this way a privileged reference frame (the
\emph{New Aether}) is introduced in the theory where a particular
simple form of the equations of motion is valid. Since Lorentz
invariance must be broken, the motion of the laboratory with respect to
the privileged frame should be detected in suitable
experiments. Moreover, modern cosmology suggests a single candidate for
the New Aether, namely, the reference system at rest with respect to
the Cosmic Microwave Background.
In this way, the possibility of accurate tests of Quantum Gravity
induced Lorentz breakdown opens.  

The rest of the paper is organized as follows: Section
\ref{ssec:LIV:Issues} is devoted to a brief discussion of the main
differences between Lorentz abiding and Lorentz violating
theories. Section \ref{sec:TestTh} discusses several of the test
theories that have been used to design and interpret tests Lorentz
invariance. In section \ref{QG:LIV} the origin of Lorentz invariance
violation in Quantum Gravity is discussed, centering on the
predictions of Loop Quantum Gravity, and in Section \ref{QG:LIV} some
of the tests carried on the theory at the tree-level are
discussed. The dramatic consequences of the inclusion of radiative
correction effects are discussed in section
\ref{sec:RadCorr}. Finally, in section \ref{sec:Concl} we state our
conclusions. 

\section{Issues of Lorentz violation}
\label{ssec:LIV:Issues}

Lorentz invariant theories have well defined properties, such as the
nonexistence of privileged reference frames. Many of these properties
are lost if Lorentz invariance does not hold and new phenomena may
arise in these conditions. In this section we discuss some of these
issues since they may be used to check the validity of Lorentz symmetry.

\subsection{Privileged reference systems}
\label{ssec:PRS}

A characteristic fact of Lorentz violating theories is that most of
them predict the existence of  \emph{privileged reference systems},
where the equations of motion take their simplest form. This is akin
to the old notion of ``luminiferous aether'', before the formulation
of special relativity one century ago, and we shall call sometimes
these privileged reference systems ``the new aether''.

That Lorentz violating theories should generically predict the
existence of privileged reference systems can be easily inferred if
propagating particles have general functions of energy as dispersion
relations. Consider for instance a photon: if its dispersion relation
does not have the Lorentz covariant form $\omega = ck$, but propagate
with an energy dependent velocity $v(E)$, such statement can be at
best valid in one specific inertial frame.  This selects a preferred
frame of reference, where a particular form of the equations of motion
is valid, and one should then be able to detect the laboratory
velocity with respect to that frame \cite{Sudarsky:2002ue}. It is this
fact that opens the possibility of detecting tiny violations of
Lorentz symmetry.

 Fortunately, we have today, in contrast with the
situation at the end of the 19th century, a rather unique choice for
that ``preferred inertial frame'': the frame where the Cosmic
Microwave Background (CMB) looks isotropic. Our velocity $\bm{w}$
with respect to that frame has already been determined to be
$w/c\approx 1.23 \times 10^{-3}$ by the measurement of the dipole term
in the CMB by COBE, for instance \cite{COBE}.  We shall usually refer
 tests of Lorentz invariance to the CMB reference system.

\subsection{``Particle'' and ``Observer'' Lorentz transformations}
\label{ssec:ActivePasive}

As it is ell known, there are two possible formulations of
Lorentz-Poincar\'e
transformations: \emph{passive} or \emph{observer} 
transformations, where coordinates are transformed in the form
\begin{equation}
  {x'} = \Lambda^{-1} (x - a) \label{def:Passive}
\end{equation}
and \emph{active} or \emph{particle}  transformations, where
 fields and states are transformed
 \begin{align}
   U(\Lambda,a) \phi(x) U^\dagger(\Lambda,a) &= \phi(\Lambda x
   + a) & U(\Lambda,a)\vert 0 \rangle &= \vert 0 \rangle
   \label{def:Active} 
 \end{align}

These two forms of the Lorentz transformation are equivalent in
Lorentz-invariant theories, but this will not be so in
Lorentz-violating theories.  In particular, particle transformations
may be limited to a given subset of Lorentz transformations, while
observer transformations, in principle, are unrestricted. These will
be interpreted as the transformations related to the laboratory $(L)$
frame, and these are important in the design and interpretation of
experiments. 

\subsection{Discrete Lorentz transformations}
\label{ssec:CPT}

Among the testable forms of Lorentz violations, the breakdown of CPT
symmetry is paramount, since it signals the breakdown of Lorentz
symmetry altogether. Indeed, the \emph{CPT Theorem}, \emph{i.e.}, the
validity of CPT symmetry, can be proved from very weak assumptions in
axiomatic field theory \cite{Streater68,Weinberg95}. The converse
proposition has been proved recently \cite{Greenberg:2002uu}.

The consequences of CPT invariance are well known: masses, magnetic
moments and charge of particles and antiparticles must be equal in
absolute value, as well as cross sections and decay rates. Any of
these properties may be used to check the validity of CPT invariance
and some of the most sensitive tests of Lorentz Invariance are based
on them.

\section{Test theories for Lorentz Invariance Violation}
\label{sec:TestTh}

Testing symmetries of nature such as Lorentz invariance is better made
through the development of \emph{test theories}: theories that
suitably generalize the symmetry being tested through the introduction
of a set of well defined parameters $C_i$. These are chosen in such a way
that the symmetry being tested is recovered for particular values of
the parameters $C^0_i$, while the other values represent a breakdown
of the symmetry. Let us examine a few of the proposed test theories of
Lorentz invariance.

\subsection{The Robertson model}
\label{ssec:Robertson}

The simplest of the proposed test theories are the Robertson model
\cite{Robertson49} and its generalization by Mansouri and Sexl
\cite{Mansouri77a,Mansouri77b,Mansouri77c}. The connection between both
references has been discussed in reference \cite{MacArthur86}. 
These are purely kinematic test theories, where the parameters are
introduced through suitable generalizations of the Lorentz
transformation. 

Consider the privileged reference system $S_0$ chosen in such a way
that Maxwell's equations are there valid. Let us introduce the metric,
valid in this rest frame
\begin{equation}
  ds^2 =  dx_0^2 -  (dx^2 + dy^2 + dz^2) \label{equ:Rest:Metric}
\end{equation}
This quantity is important, since it has the following properties

\begin{enumerate}
  \item A time interval at a given point $\bm{x}$ is expressed as
    \begin{equation}
      dt = \frac{ds}{c} \qquad (d\bm{r} = 0)
    \end{equation}
  \item The distance between two points $\bm{x}_1,\bm{x}_2$ is
  expressed as
  \begin{equation}
    dl = +\sqrt{-ds^2} \qquad (dt = 0)
  \end{equation}
  \item Light rays are along the normal to the spheres
    \begin{equation}
      ds^2 = 0
    \end{equation}
\end{enumerate}

Now consider an inertial reference system $S'$, moving with respect to
$S$ with velocity $\bm{w}$. The Robertson transformation between
both reference systems,  with a suitable choice of axis, is defined as
\begin{equation}
  \begin{split}
    x &= A x' + B x'_0\\
    x_0 &= B' x_0' + A' x'\\
    y &= C y'\\
    z &= C z'
  \end{split}\label{def:Rob:Trans}
\end{equation}
with $A, B, C$ arbitrary functions of the relative velocity $w$. These
are, of course, ``observer'' transformations. 

Imposing  Einstein's synchronization on $S'$ the conditions
\begin{align}
  A' &= \frac{w}{c} A & B' &= \frac{w}{c} B
\end{align}
are obtained. Finally, the metric in $S'$ results
\begin{equation}
  ds^2 = g_0^2 {dx_0'}^2 - g_1^2 {dx'}^2 -  g_2^2 ({dy'}^2 + {dz'}^2)
  \label{equ:Robert:Metric:Sp} 
\end{equation}
where
\begin{equation}
  \begin{split}
    g_0(w) &= \sqrt{1 - \frac{w^2}{c^2}} B\\
    g_1(w) &= \sqrt{1 - \frac{w^2}{c^2}} A\\
    g_2(w) &= C
  \end{split}\label{def:g0,g1,g2}
\end{equation}

Thus, these tests theories are equivalent to the proposal of a metric of the
form
\begin{equation}
  ds^2 = g_0^2 dx_0^2 - g_1^2 dx^2 - g_2^2 (dy^2 + dz^2)
  \label{equ:R-Metric} 
\end{equation}
for space-time, where $g_i(V)$ are arbitrary functions of the velocity
of the local reference system with respect to a privileged inertial
frame. Lorentz symmetry is recovered for the particular case $g_i(V) =
1$. For low velocity experiments, it is usual to parametrize these
functions in terms of the \emph{Mansouri-Sexl parameters}
\cite{Mansouri77a} 
\begin{gather}
  g_0(V) = 1 + \left(\frac{1}{2} - \alpha\right)\frac{V^2}{c^2} +
  \dots\\
  \frac{g_1}{g_0} = 1 + \left(\alpha - \beta\right)\frac{V^2}{c^2} +
  \dots\\
  \frac{g_2}{g_1} = 1 + \left(\delta - \beta\right)\frac{V^2}{c^2} +
  \dots
\end{gather}
which have the values
\begin{align}
  \alpha &= \frac{1}{2} & \beta &= \frac{1}{2} & \delta &= 0
\end{align}
if Lorentz invariance is valid.

Robertson \cite{Robertson49} showed that the classical
experiments of Michelson-Morley \cite{MM87}, Kennedy-Thorndike
\cite{KT32} and Yves-Stillwell \cite{YS38} restricted the functions
$g_i$ to a small neighborhood of 1. We shall discuss modern limits on $g_i$
later on. 

\subsection{The Dispersion Relations models}
\label{ssec:AC:Model}

A set of simple tests theories for testing Lorentz violation has been
proposed by Amelino-Camelia \cite{Amelino-Camelia:2004ht} (See also
\cite{Konopka:2002tt}). These are kinematic models, but applied to the
dynamical dispersion relations they introduce a bridge between
kinematics an dynamics.

The dispersion relation for a low-energy particle (\emph{i.e.} small
with respect to the Planck scale $E_P = \unit[1.2\times10^{19}]{GeV}$)
in a Lorentz violating theory, can be written in the form
\begin{equation}
  m^2 \simeq E^2 - \bm{p}^2 + \eta \bm{p}^2
  \left(\frac{E}{E_P}\right)^n +
  O\left(\frac{E}{E_P}\right)^{n+3} \label{def:AC:Test}
\end{equation}
where $\eta$ and $n$ are free parameters. The exponent $n$ is
characteristic of the magnitude of the effects expected. On the other
hand, for Lorentz violating effect generated in quantum gravity, one
expects $\eta \sim O(1)$. 
The case $n=1$ is specially interesting, since it arises in many
contexts such as low energy limits of Loop Quantum Gravity.%
{It should be noted, however, that a photon dispersion
  relation of the form \eqref{def:AC:Test} with $n$ odd is forbidden
  by causality. The detection of such a correction would signal not
  only the breakdown of Lorentz invariance but of causality
  relations.}

To close the test theory, two prescriptions must be added to the above
dispersion relation
\begin{enumerate}
  \item The law of energy-momentum conservation is valid
    \begin{align}
      E_1 + E_2 &= E'_1 + E'_2 & \bm{p}_1 + \bm{p}_2 &= \bm{p}'_1 +
      \bm{p}'_2 \label{equ:AC:EpCons}
    \end{align}
  \item The velocity of the particle is computed from the Lorentzian
  expression
  \begin{equation}
    \bm{v} = \frac{dE}{d\bm{p}} \label{equ:AC:v}
  \end{equation}
\end{enumerate}

These simple prescriptions suffice for analyzing many
Lorentz-violating phenomena, specially those related to threshold
modifications by Quantum Gravity effects.

The set of equations \eqref{def:AC:Test}, \eqref{equ:AC:EpCons} and
\eqref{equ:AC:v} are the basis of \emph{threshold analysis}: the very
powerful tool to discuss many astrophysical phenomena in the presence
of Lorentz violation \cite{Alfaro:2002ya,Lehnert:2003ue,%
Amelino-Camelia:2004ht,Amelino-Camelia:2004hm}.

A variation of the above dispersion relation is obtained assuming that
different polarization states of the particle have different
propagation velocities
\begin{equation}
  m^2 = E^2 - \bm{p}^2 \pm \theta_i \bm{p}^2\left(\frac{E}{E_P}\right)
  \label{equ:AC:Pol} 
\end{equation}
where the sign depends on the helicity of the particle. This type of
test theories should be analyzed in the context of low energy field
theories. 

\subsection{The $TH\epsilon\mu$ model}
\label{ssec:THem}

A \emph{dynamical} test theory for local Lorentz invariance was
introduced in reference \cite{LightLee73} as a test theory for the
validity of Einstein's Equivalence Principle. In this model, the
breakdown of Local Lorentz Invariance comes from the structure of the
Lagrangian of the system. 

Let us consider a system of classical particles interacting through
the electromagnetic field in a background spherically symmetric
gravitational field $U$. The action of the system is 
\begin{equation}
  \begin{split}
  S_{TH\epsilon\mu} =& - \sum_a m_{0a} \int \sqrt{T - H v_a^2} dt \\
    &+ \sum_a e_a \int A_\mu(x^\nu_a) v^\mu_a dt\\
    &+ \frac{1}{8\pi} \int \left(\epsilon \bm{E}^2 -
    \frac{\bm{B}^2}{\mu}\right) 
  \end{split}\label{equ:S-THem}
\end{equation}
where $\hbar=c=1$, $m_a, e_a$ are the mass and charge of the $a$-th
particle, $x^\mu_a, v^\mu_a$ its world-line and velocity and $\bm{E},
\bm{B}$ the electric and magnetic fields. $T, H, \epsilon, \mu$ are
arbitrary functions of the spherically symmetrical gravitational field
$U = GM/r$. 

To see how a Lorentz invariance violation is generated in this test
theory, let us pass to a freely falling reference frame. Consider a
given point in space-time $x^\mu_0 = (t_0, \bm{r}_0)$, and a local
reference frame with origin in $x^\mu_0$. If $\bm{g} = \bm{\nabla}
U$ is the local acceleration of gravity, the transformation equations
to the freely falling reference frame, correct to first order in the
small quantities $\bm{g} t$ and $\bm{g\cdot x}$ are

  \begin{gather}
  \hat{t} = \sqrt{T_0}t\left(1 + \frac{T'_0}{2T_0} \bm{g\cdot
  x}\right)\\ 
  \hat{x} = \sqrt{H_0}\left[\bm{x} + \frac{T'_0}{4H_0}\bm{g}t^2 +
  \frac{H'_0}{4H_0}(2\bm{x}(\bm{ g\cdot x}) - \bm{g} x^2)\right]
  \end{gather}

The transformed action, keeping only the lowest order terms, is
\begin{equation}
  \begin{split}
    \hat{S}_{TH\epsilon\mu} =& - \sum_a m_{0a} \int \sqrt{1 -
    \hat{v}^2} d\hat{t}\\
    &+ \sum_a e_a \int \hat{A}_\mu \hat{v}^\mu_a d\hat{t}\\
    &+ \frac{1}{8\pi} \epsilon_0 \sqrt{\frac{T_0}{H_0}}
    \int\left(\hat{E}^2 -
    \frac{H_0}{T_0\epsilon_0\mu_0}\hat{B}^2\right) d^4\hat{x}
  \end{split}\label{equ:S-THem:FF}
\end{equation}
where the ``hatted'' quantities are referred to the local reference
system. The first two terms in \eqref{equ:S-THem:FF} are locally
Lorentz invariant, but the third one breaks the symmetry unless
\begin{align}
  \frac{H_0}{T_0\epsilon_0\mu_0} &= 1 & \epsilon_0
  \sqrt{\frac{T_0}{H_0}} &= 1 \label{equ:THem:Conds}
\end{align}
which are the conditions for the validity of Local Lorentz
Invariance. The first one implies that the limiting speed for
massive particles is equal to the propagation speed for
electromagnetic signals
\begin{equation}
  c_L^2 = \frac{T_0}{H_0} = \frac{1}{\epsilon_0\mu_0} = c^2_{em}
  \label{equ:THem:Cond1} 
\end{equation}

The second one, which is related to the validity of the Weak
Equivalence Principle, can be recast as position invariance of the
electric charge. 

The $TH\epsilon\mu$ model has been generalized to the Standard Model
\cite{GMSM} where a set of conditions similar to
\eqref{equ:THem:Conds} should hold for the validity of Lorentz
symmetry.

\subsection{The Kosteleck\'y Model}
\label{ssec:K-Model}

The most general treatment of Lorentz Invariance violations within the
Standard Model is the model developed by Kosteleck\'y and coworkers
\cite{Colladay:1996iz,Colladay:1998fq,Kostelecky:1999rh,Kostelecky:2003fs},
which we shall simply call the Kosteleck\'y Model (KM).

The KM Lagrangian density is based on a careful inclusion of all terms
of dimension 4 that break Lorentz invariance for the particles in the
Standard Model. Thus, the Dirac particle modified Lagrangian takes the
form

  \begin{equation}
    \mathcal{L}_D = \frac{1}{2}i\bar{\psi}\Gamma^\mu\partial_\mu\psi -
    \bar{\psi}M\psi  \label{equ:KM:Dirac}
  \end{equation}
  where the $\Gamma$ and $M$ operators are
  \begin{gather}
    \Gamma^\mu = \gamma^\mu + c^{\mu\nu}\gamma_\nu +
    d^{\mu\nu}\gamma_5\gamma_\nu + e^\mu + if^\mu\gamma_5 +
    \frac{1}{2}ig^{\lambda\nu\mu}\sigma_{\lambda\nu} \label{equ:KM:Gamma}\\
    M = m + a_\mu\gamma^\mu + b_\mu\gamma_5\gamma^\mu +
    \frac{1}{2}H^{\mu\nu}\sigma_{\mu\nu} \label{equ:KM:Mop}
  \end{gather}

These operators have well defined transformation properties under CPT
symmetry. In particular, the $a_\mu$ and $b_\mu$ terms are CPT-odd,
making the masses of particle and antiparticle different. 

There are corresponding modifications for the gauge field and Higgs
Lagrangians. For the particular case of the photon Lagrangian, these
modifications take the form
  \begin{equation}
     \mathcal{L}_{em} = - \frac{1}{4}(k_F)_{\kappa\lambda\mu\nu}
  F^{\kappa\lambda}F^{\mu\nu} + \frac{1}{2} (k_{AF})^\kappa
  \epsilon_{\kappa\lambda\mu\nu} A^\lambda F^{\mu\nu} 
  \end{equation}
where the first term is CPT-even while the second is CPT-odd.

The CPT-even tensor $k_F$ has 19 independent components, which can be
grouped into a set of 10 P-odd components (responsible for vacuum
birefringence phenomena) and a set of 9 P-even components, which
describe breakdown of Lorentz boost invariance
\cite{Kostelecky:2001mb,Kostelecky:2002hh,mueller03}.  

KM is an extremely general test theory for Lorentz violation: it
includes all dimension  4- terms and thus it is ``closed'' under
Lorentz transformations. The physical properties of the model have been
extensively studied
\cite{Colladay:1996iz,Kostelecky:1999zh,Kostelecky:2000mm,%
Kostelecky:2001jc,Kostelecky:2003fs,Bailey:2004na,Montemayor:2004uy} 
and it has been successfully used to design and analyze
\cite{Bluhm:1997ci,Bluhm:1998rk,Bluhm:1999ev,Bluhm:2001rw,Bluhm:2001ms,%
Bluhm:2003un,Bluhm:2003ne,Bluhm:2004tm} 
experiments on Lorentz violation. 

The nonrelativistic limit of the generalized Dirac operator is
specially interesting since same of the most accurate test of Lorentz
invariance have been carried in that r\'egime. Although the Lorentz
violating terms are small and can be treated perturbatively, it is not
straightforward to carry out the nonrelativistic limit because several
of the small terms have nontrivial time derivatives. A generalization
of the Foldy-Wouthuysen transformation developed n reference
\cite{Kostelecky:1999zh} can be used to obtain it. Neglecting terms of
order $(P/m)^3$ the following Hamiltonian is obtained
\begin{equation}
  \begin{split}
    H_{NR} =& m + \frac{p^2}{2m}\\
           &+ [a_0 + m(c_{00} + e_0)] + 
              [-b_j + \frac{1}{2} \epsilon_{jkl} H_{kl} + m(d_{j0} -
                \epsilon_{jkl} g_{kl0})] \sigma^j\\
	   &+ [-a_j + m(c_{j0} + c_{0j} + e_j)]\frac{p^j}{m}\\
	   &+ [b_0\delta_{kj} - \epsilon_{jkl} H_{l0} - m(d_kj +
              d_{00}\delta_{kj} + \frac{1}{2}\epsilon_{klm}g_{mlj} -
              \epsilon_{jkm} g_{m00})] \frac{p^j}{m}\sigma^k\\
	   &- m(c_{jk} + \frac{1}{2}c_{00})\frac{p^jp^k}{m^2}\\
	   &+ \{[m(d_{j0} + d_{0j}) - \frac{1}{2}(b_j + md_j +
           \frac{1}{2} m \epsilon_{jmn}g_{mn0} +
           \frac{1}{2}\epsilon_{jmn} H_{mn})] \delta_{kl}\\
	      &+ \frac{1}{2}(b_l + \frac{1}{2}m \epsilon_{lmn}g_{mn0})
           \delta_{jk} - m \epsilon_{jlm}(g_{m0k} + g_{ml0})\}
           \frac{p^jp^k}{m^2} \sigma^l
  \end{split}\label{equ:KM:H-NR}
\end{equation}

The first line of \eqref{equ:KM:H-NR} corresponds to the usual NR
Hamiltonian for a free particle. The rest of the terms describe
Lorentz violating phenomena of different types. We shall discuss some
of these phenomena in the following pages.

\subsection{Other models of Lorentz violation}
\label{ssec:OtherLIV}

As a last example of test theories for Lorentz violation, let us
mention the Myers-Pospelov model \cite{Myers:2003fd,Myers:2004ge}. In
this model, terms of dimension five are added to the Lagrangian of the
standard model satisfying the following criteria
\begin{enumerate}
  \item The terms must be quadratic in the same field.
  \item They should have one more derivative than the kinetic energy
  term.
  \item They must respect gauge invariance.
  \item They must be Lorentz invariant except for the appearance of a
  unit vector $n\cdot n = 1$ parameterizing a privileged direction in
  spacetime. 
  \item The terms should not reduce to lower dimension terms through
  the equations of motion.
  \item They should not reduce to a total derivative.
\end{enumerate}

The simplest correction satisfying the above criteria are 
\begin{gather}
  \delta\mathcal{L}_S = i\kappa\ell_P \phi(n\cdot\partial)^3\phi
  \label{equ:MyP-scalar} \\
  \delta\mathcal{L}_\gamma = \xi\ell_P n^\mu F_{\mu\alpha}
  (n\cdot\partial) n_\nu \tilde{F}^{\nu\alpha} \label{equ:MyP-gamma}\\
  \delta\mathcal{L}_f = \ell_P \tilde{\psi} \left(\eta_1\not{n} +
  \eta_2\not{n}\gamma_5\right) (n\cdot\partial)^2 \psi
  \label{equ:MyP-fermion} 
\end{gather}

In these equations $\xi, \eta_1, \eta_2$ are free parameters and 
\begin{equation}
  \tilde{F}_{\mu\nu} = \frac{1}{2} \epsilon_{\mu\nu\alpha\beta}
  F^{\alpha\beta} 
\end{equation}
is the dual of the electromagnetic field tensor.  

These terms are CPT-odd and thus induce a host of Lorentz breaking
phenomena. The model can be applied to the tree level, but radiative
corrections will generate terms of dimension 2 or 3 badly breaking
Lorentz invariance \cite{Perez:2003un,Collins:2004bp}. These issues
will be treated later on.

Last but not least I would like to mention \emph{double special
relativity} as a possible test theory, not of Lorentz violation but of
new physics near the Planck scale
\cite{Magueijo:2001cr,Amelino-Camelia:2002wr}.  Rather than a
violation of Lorentz invariance, double special relativity is a
nonlinear realization of the Lorentz group, that exhibits both a
limiting velocity (the speed of light $c$) and a limiting energy (he
Planck energy $E_P$). This beautiful theory falls beyond the scope of
this article, which is concerned with ``true'' violations of Lorentz
invariance. For a discussion on double special relativity, see for
instance \cite{Amelino-Camelia:2003ex}.

\section{Quantum Gravity as a source of Lorentz violation}
\label{QG:LIV}

Quantum Gravity has been proposed as a source of Lorentz violations
\cite{Kostelecky:1988zi,Amelino-Camelia:1997gz,Gambini:1998it,%
Carmona:2002iv,%
GrootNibbelink:2004za} by many authors. The rationale behind the
proposal is that quantum gravity effects will make space-time granular
at a scale near the Planck scale $\ell_P$. This in turn will change
the dispersion relations of elementary particles, breaking in general
Lorentz invariance.

As it has been mentions, such modifications have been found within
the two currently most popular approaches to quantum gravity, namely
\emph{Loop Quantum Gravity} and \emph{String theory}. Let us discuss
briefly how these phenomena arise, taking Loop Quantum Gravity as an
example. The analysis of the resulting Lorentz violations can be
carried out in the framework of the test theory developed by
Kosteleck\'y and coworkers.

\subsection{A reminder of Loop Quantum Gravity}
\label{ssec:LQGPrim}

Loop Quantum Gravity is a theory of canonical quantization of the
gravitational field. References
\cite{Thiemann:2002nj,Perez:2004hj,Smolin:2004sx} are introductions in
different levels, while \cite{Nicolai:2005mc} is a critical review on
the subject. 

The details of the theory are complicated, so in this note we shall only
give some qualitative description.

\begin{figure}[t]
\begin{center}
\begin{minipage}[h]{80mm}
\epsfig{file=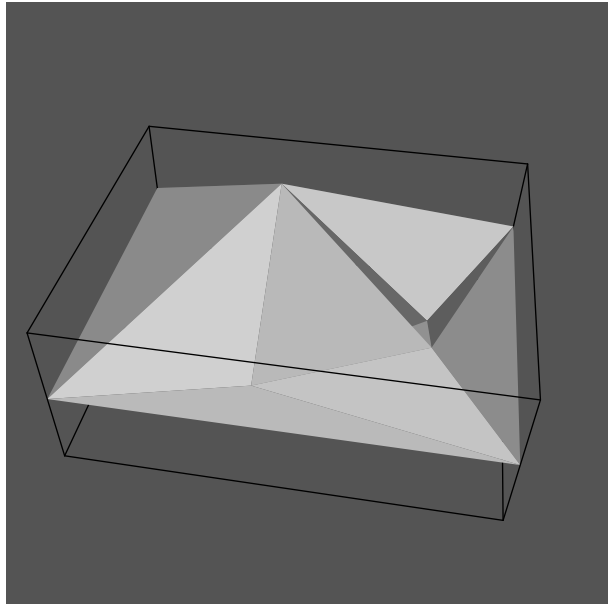, width=80mm}
\end{minipage}
\caption{A discretized state of the gravitational field.}
\label{fig:Weave} 
\end{center}
\end{figure}

As it is well-known, gravitation has gauge invariance under
diffeomorphisms and thus physical states of the gravitational field
must satisfy the \emph{momentum constraint} and the \emph{Hamiltonian
  constraint}. The strategy of Loop Quantum Gravity is to find a
suitable way of dealing with these constraints.
\begin{enumerate}
  \item In order to perform the quantization, a suitable set of
  gravitational canonical variables must be introduced. These are
  cunningly contrived modifications of the Ashtekar variables
  \cite{Ashtekar:1986yd,Ashtekar:1987gu}.
  \item A regularization is introduced in the theory by breaking
  space-time in a lattice-like structure, at a scale of the order of
  $\ell_P$ (Figure \ref{fig:Weave}). With this regularization, all
  physical quantities became well defined, although gauge
  dependent. \label{step:Regul}
  \item In terms of these variables,a suitable auxiliary Hilbert space
  ${\mathcal{H}_{\mathrm{aux}}}$ is built
  \cite{Thiemann:2002nj}. However, because of the invariance of the
  theory under diffeomorphisms, not all of the states in
  ${\mathcal{H}_{\mathrm{aux}}}$ represent physical states of the
  gravitational field. 
  \item The physical states are chosen so that the Wilson loops built
  from the canonical variables are well defined. This is the crucial
  step, from which the theory can be developed. This representation,
  originally introduced for the electromagnetic field
  \cite{Gambini:1980wm,Gambini:1980yz,diBartolo:1983pt} allows the
  definition of well defined physical quantities. 
  \item In principle, the regularization introduced in step
  \ref{step:Regul} can be eliminated shrinking the scale of the
  lattice to zero. However, it can be shown that the volume element
  operator acts like a regulator, yielding well-defined quantities in
  this limit.
\end{enumerate}

Although the technical details are very complex, the above steps have
been carried to success (see, however,
Ref. \cite{Nicolai:2005mc}). Among the most interesting results
obtained, let us mention
\begin{enumerate}
  \item Geometrical quantities, such as areas and volumes, are
  quantized s multiples of $\ell_P^2$ (or $\ell_P^3$), even in the
  limit of the auxiliary lattice size going to zero
  \cite{Thiemann:2002nj,Perez:2004hj}. 
  \item Other field theories can be included in the scheme. The
  lattice-like discretization provides a suitable regularization of
  the field quantities \cite{Thiemann:2002nj}. 
  \item Solutions for highly symmetrical states, such as spherically
  symmetrical or cosmological ones, can be found by solving additional
  constraints. For instance, a beautiful solution for a
  Robertson-Walker universe has been found, free of singularities (For
  a review, see \cite{Bojowald:2003uh,Bojowald:2004ax}).
\end{enumerate}

\subsection{Lorentz Invariance Violation in Loop Quantum Gravity}
\label{ssec:LIV-LQG}

Let us examine the origin of Lorentz invariance violation in Loop
Quantum Gravity with an example: the lowest order corrections to the
Maxwell Hamiltonian \cite{Gambini:1998it}. The latter can be written
in the form
\begin{equation}
  H_{\rm M} = \frac{1}{2\alpha} \int d^3x \frac{g_{ab}}{g}
  \left(\tilde{E}^a\tilde{E}^b + \tilde{B}^a\tilde{B}^b\right)
  \label{equ:H-M} 
\end{equation}
where the tildes denote that the fields are tensor densities in the
canonical framework, $g$ is the determinant of the metric and $\alpha$
is the fine structure constant. We wish
to average \eqref{equ:H-M} over a semiclassical state
\ket{g,\CL}. Indexes $a,b$ are purely spatial.

There is currently no clear characterization in Loop Quantum Gravity
of a semiclassical state. Following \cite{Gambini:1998it} we shall use
a \emph{weave}: this is a state characterized by a semiclassical
length scale \CL\ such that
\begin{equation}
  \ell_P \ll \CL \ll \lambda \label{equ:lP-CL-lambda} 
\end{equation}

\begin{figure}[t]
\begin{center}
\begin{minipage}[h]{80mm}
\epsfig{file=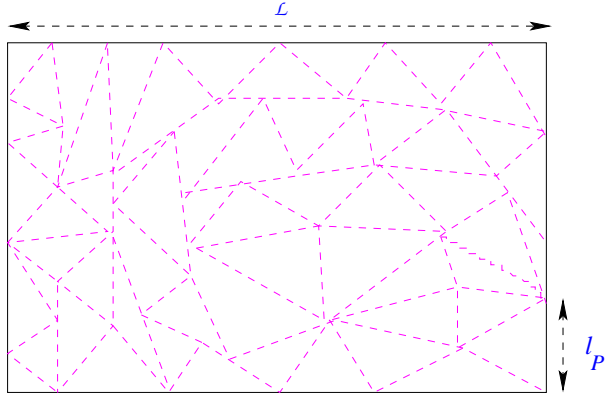, width=80mm}
\end{minipage}
\caption{A semiclassical weave state}
\label{fig:SemSt}
\end{center}
\end{figure}

Intuitively, a weave can be thought as a big carpet, of size \CL,
weaved of tiny pieces of scale $\ell_P$ (Fig. \ref{fig:SemSt}). The
characteristic size \CL\ should be smaller than any ``macroscopic''
characteristic length $\lambda$ such that spacetime can be considered
continuous. Thus, weaves are states where whose scale \CL\ marks the
transition from quantum (discrete) spacetime to classical one. 

In the following, we shall assume that the weave state \ket{g,\CL}\
correctly describes the \emph{privileged} reference frame $S$.

Consider the averages of the Hamiltonian \eqref{equ:H-M} in the weave
state \ket{g,\CL}
\begin{equation}
  H_{\rm ef} = \braket{g,\CL}{H_{\rm M}}{g,\CL}
\end{equation}
The average is performed for fixed values of the electromagnetic field
operators $\bm{E},\bm{B}$ which, for all practical purposes, can be
considered classical fields. That is, the average is made on
the fluctuations of the quantum gravitational field. We shall also
assume that the classical geometry in the privileged reference frame
is flat euclidean and so we expect
\begin{subequations}
  \begin{gather}
   \braket{g,\CL}{{g_{ab}}{g}^{-1}}{g,\CL} = \delta_{ab} +
  O\left(\frac{\ell_P}{\CL}\right) \label{equ:ValM(gab)}\\
   \braket{g,\CL}{A_{ia}}{g,\CL} = 0 +
  \frac{1}{\CL}\left(\frac{\ell_P}{\CL}\right)^\Upsilon
  \label{equ:ValM(Aia)} 
  \end{gather}
\end{subequations}
where $A$ is the Ashtekar variable operator and $\Upsilon$ is a real
number parameter.

Let us now introduce the Thiemann factorization of the metric operator
\begin{equation}
  {g_{ab}}{g}^{-1} = \hat{w}_a \hat{w}_b 
\end{equation}
where the operators $\hat{w}$ are finite and have nonzero
contributions only at the vertexes $v_i$. Introducing a point
splitting regulator $\Delta(x,y) \to \delta(x-y)$ when the
triangulation scale goes to zero, the Maxwell Hamiltonian can be
written as a sum over the vertexes in a weave
\begin{equation}
  H_{\rm ef} = \frac{1}{2\alpha} \sum_{v_i,v_j}
  \braket{g,\CL}{\hat{w}(v_i)\hat{w}(v_j)}{g,\CL}
  \left[E^a(v_i)E^b(v_j) + B^a(v_i)B^b(v_j)\right]  
\end{equation}

The electromagnetic field, being quasiclassical, is slowly varying
across the weave and can be expanded in a Taylor series about some
point $x_0$ 
\begin{equation}
  E^a(v_i) = E^a(x_0) + (v_i - x_0)_c \partial^c E^a(x_0)
\end{equation}
and after a simple calculation the low energy Maxwell Hamiltonian
takes the form
\begin{equation}
  H_{\rm ef} = \frac{1}{2\alpha} \int d^3x \left[\left(\bm{E}^2 + \bm{B}^2\right)
  + 2 t_{abc}\left(E^a\partial^cE^b + B^a\partial^cB^b\right)\right]
\end{equation}
where the 3-tensor 
\begin{equation}
  t_{abc} =
  \frac{1}{2}\sum_{v_i,v_j}\braket{g,\CL}{\hat{w}(v_i)\hat{w}(v_j)}{g,\CL}
  (v_i - x_0)_c 
\end{equation}

Now, let us consider the electromagnetic field in the
\emph{privileged} reference frame $S$. We expect that the
electromagnetic field will behave isotropically in this particular
reference frame and so we shall write the averages as isotropic
tensors in $S$: that is, tensors built from the metric tensor
$\delta_{ab}$ or the Levi-Civita tensor $\epsilon_{abc}$. Thus the
3-tensor $t_{abc}$ must have the form
\begin{equation}
  t_{abc} = \theta_{GP} \ell_P \epsilon_{abc}
\end{equation}
the last factor coming from the fact that a nonzero value of $t_{abc}$
can be expected only at scales near the Planck length, where the
distributions of vertexes is very anisotropic.

As a final result, we get the \emph{Gambini-Pullin Hamiltonian}
\begin{equation}
  H_{GP} = \frac{1}{2\alpha} \int d^3x \left[\left(\bm{E}^2 +
  \bm{B}^2\right) + 2\theta_{GP} \ell_P \left(\bm{E\cdot\nabla\times E}
  + \bm{B\cdot\nabla\times B}\right)\right] \label{equ:H-GP}
\end{equation}

The second term is parity odd and originates birefringence in vacuum
propagation. This happens because we have implicitly assumed that
weaves can break parity, although in the privileged system rotational
invariance is preserved. 

The above informal method was developed in a much more rigorous way in
reference \cite{Alfaro:2001rb}, where a detailed exposition of the
used techniques is given, and higher order corrections in the
small parameter $\ell_P/\CL$ were found. These higher order
corrections depend on the  $\Upsilon$ and \CL\ parameters and have a
rather complex structure. 

The method was also applied to the Dirac Hamiltonian
\cite{Alfaro:1999wd,Alfaro:2002xz}. These latter corrections are
particularly important, since they can be submitted to strict
experimental tests.

The modified Dirac equation, in the privileged reference frame, takes
the form
\begin{equation}
    \left[ i\gamma^{\mu }\partial_{\mu } + \Theta_1 
    m\ell_{P} i\bm{\gamma\cdot\nabla} -{K}/{2} \gamma_{5} \gamma^{0}
    - m\left( \alpha -i\Theta_2 \ell_{P} {\bm{\Sigma\cdot \nabla}}
    \right) \right] \Psi = 0
    \label{DIREQ}
\end{equation}
where the spin operator is {$\Sigma ^{k}=(i/2)\epsilon
_{klm}\gamma ^{l}\gamma ^{m}$}.  Here 
\begin{align}
 {K}&=m{\Theta_4}m\ell_{P}, & \alpha &=\left( 1+{\Theta_3}m\ell
_{P}\right) 
\end{align}
and 
$\Theta_1,\Theta_2,\Theta_3,\Theta_4$ are constants of order
one. Besides we have assumed {${\cal L}$ =$1/m$},
where {$m$} is typically the particle mass. 

 The term
{$m\left( 1+{\Theta_3\;}m\ell _{P}\right) $} can be
interpreted as a renormalization of the mass whose physical value is
taken to be {$M=m\left( 1+{\Theta_3\;}m\ell
_{P}\right) $}. The other terms, however cannot be eliminated by a
simple transformation, since they break discrete symmetries. Indeed,
the first two correction terms are CPT odd, while the last one is CP
odd. Thus, they represent new physical effects induced by Quantum
Gravity. 

\subsection{Transforming into the Laboratory System}
\label{ssec:PtoL}

Although many phenomena can be conveniently analyzed in the Privileged
reference system $S$, many others are manifest in the Laboratory
reference system $S'$, moving with four velocity $W^\mu$ with respect
to $S$. This transformation can be accomplished in a very elegant way
using the Lagrangian density, which should be a scalar density
\cite{Sudarsky:2002ue}. As an example, consider transforming
\eqref{DIREQ} to the lab system.

We shall write \eqref{DIREQ} in a formally covariant form using
$W^\mu$ and projectors constructed from it. In this way, we find
\begin{equation}
  \begin{split}
    \mathcal{L} =& \frac{1}{2}\bar{\Psi} i\gamma^\mu\partial_\mu \Psi
    - \frac{1}{2}\bar{\Psi} M \Psi \\
    &+ \frac{1}{2} i(\Theta_1M\ell_P) \bar{\Psi} \gamma_\mu
    \left(g^{\mu\nu} - W^\mu W^\nu\right)\partial_\nu \Psi \\
    &+ \frac{1}{4} (\Theta_2M\ell_P) \bar{\Psi}
    \epsilon_{\mu\nu\alpha\beta} W^\mu \gamma^\nu\gamma^\alpha
    \partial^\beta \Psi\\
    &- \frac{1}{4} (\Theta_4M\ell_P) MW^\mu \bar{\Psi}
    \gamma_5\gamma_\mu \Psi
  \end{split}\label{equ:L-Lab}
\end{equation}

In this way, the Dirac particle Lagrangian takes the form of the
Kosteleck\'y Lagrangian, with
\begin{gather}
  a_\mu = H_{\mu\nu} = d_{\mu\nu} = e_\mu = f_\mu = 0\\
  c_{\mu\nu} = \Theta_1M\ell_P\left(g^{\mu\nu} - W^\mu W^\nu\right)\\
  g_{\alpha\beta\gamma} = \Theta_2M\ell_P
  \epsilon_{\mu\alpha\beta\gamma} W^\mu \\
  b_\mu = \frac{1}{2} \Theta_4 M^2 \ell_P W^\mu
\end{gather}
and from these expressions the low energy Hamiltonian for the fermion
can be found in the form
\begin{eqnarray} 
\tilde{H}&=&  Mc^2(1+\Theta_1\, M\ell_P\,\left({\bm{w}}/{c} \right) 
^{2}) \nonumber\\
& & + \left(1+2\,\Theta_1M\ell
_{P}\left(1+\frac{5}{6}\left({\bm{w}}/{c} \right)  
^{2}\right)\right)\left(\frac{p^{2}}{2M}+g\,\mu 
\,{\bm{s}}%
\cdot {\bm{B}}\right) +\nonumber\\ 
&& + \left(\Theta_2+\frac{1}{2}\Theta_4 \right)M\ell _{P}\left[\left(2Mc^2 
-\frac{2p^{2}}{3M}\right)\,{\bm{s}}%
\cdot \frac{\bm{w}}{c}+\frac{1}{M}\,{\bm{s}}\cdot {Q}_{P}\cdot
\frac{\bm{w}}{c}\right]  \nonumber\\ 
&& +\Theta_1M\ell _{P} %
\left[ \frac{{\bm{w}}\cdot {Q}_{P}\cdot {\bm{w}}}{Mc^2}\right], 
\label{equ:B6} 
\end{eqnarray} 
where $g$ is the standard gyromagnetic factor and ${Q}_{P}$ represents
the momentum quadrupole tensor, with components $Q_{Pij} =
\left\langle p_ip_j - 1/3p^2 \delta_{ij}\right\rangle$. 

The first two lines of \eqref{equ:B6} represent the Schr\"odinger-Pauli
Hamiltonian, with corrections to the rest and inertial mass, coming
from the motion with respect to the ``New Aether''. The third and
fourth lines represent an ``Aether wind'' blowing on the spin and and
an anisotropy of the inertial mass of the particle, which could be
tested in Hughes-Drever experiments. 

\subsection{Other models of Lorentz violation by Quantum Gravity}
\label{ssec:OtherGQ-LIV}

A similar treatment can be applied to other models of Lorentz
violation generated by Quantum Gravity. As an example let us briefly
discuss the so called Liouville approach to Non-critical String
Theory.  In this scheme our universe is
identified with a 4 dimensional D-brane, which will naturally contains
a type of topological defect called a D-particle. These will interact
with ordinary particles, represented by strings in the corresponding
mode, by elastic scattering, which would produce a recoil of the
D-particle. This recoil will produce a local disturbance of the
space-time geometry which will in turn affect the propagation of the
particle \cite{Ellis:1999uh,Ellis:1999sf,Ellis:1999sd}.

 We take this part of the analysis
directly from Ref. \cite{Ellis:2002bu}, starting with their modified Dirac
equation
\begin{eqnarray}
\left[{\gamma}^{\mu }(i\partial _{\mu }-eA_{\mu 
})-v^{i}{\gamma}
^{0}(i\partial _{i}-eA_{i})-m\right] \Psi &=&0.
\label{DIREQ2}
\end{eqnarray}
Here $\gamma^\mu$ are the standard flat-space gamma matrices and $v^i$
is the D-particle recoil velocity in the CMB frame, where it is
initially at rest. The above modified Dirac equation preserves gauge
invariance due to the standard minimal coupling. In the following we
set $A_\mu=0$. We will be concerned with a
perturbative expansion in the small parameter $v^{i}$. 

The above equation can be interpreted as describing the motion of a
fermion in an effective metric
\begin{equation}
   p^\mu\,G_{\mu\nu}\,p^\nu=E^2-2E\bm{v\cdot p}- \bm{p}{}^2, \quad
   p^{\mu}=(E,  \bm{p}), \quad | \bm{v}|\approx 
\frac{\mid \bm{p} \mid }{M}.
\label{metric}
\end{equation}

The analysis of the above equation requires some care to define
correctly the D-particle recoil velocity in the nonrelativistic limit
\cite{Sudarsky:2002zy}.  Passing to the laboratory frame the following
form of the equation is found
\begin{equation}
 \left[i\gamma ^{\mu }   \partial _{\mu } - i\gamma ^{\rho
}\left(W_{\rho} V^{\nu} \right) \partial _{\nu } -m \right]\Psi =0,
                            \label{eq:DiracMod}
\end{equation}
which can be put in the  standard form of \cite{Kostelecky:1999zh}. The
identification of  the  corresponding Kosteleck\'y parameters leads to  
\begin{eqnarray}
\label{DEFLANE}
a_{\mu } &=&b_{\mu }=0=H_{\mu \nu }, \qquad
g_{\lambda \mu \nu } = 0=d_{\mu \nu }, \qquad
c_{\mu \nu } = - W_{\mu }V_{\nu }.
\end{eqnarray}
Let us note that $c_{\mu\nu}$ is not symmetrical in this case. 

Non commutative field theory is another important model of Lorentz
violating field theory. We shall discuss issues connected with these
theories in Section \ref{sec:RadCorr}.

\section{Testing Lorentz Invariance Violation at the tree level}
\label{sec:LIVTest}

The results collected in Section \ref{QG:LIV} can be tested with
sensitive well defined experiments if one neglects the effects of
radiative corrections. In this section we shall examine these tests at
the ``tree level'', leaving a discussion of radiative corrections for
the next one.

There is a host of phenomena which can be used to detect Lorentz
Invariance violation. Among them, let us mention
\begin{enumerate}
  \item Breakdown of local rotational symmetry (``Aether wind
  effects'').
  \item Breakdown of Lorentz Boost Invariance (``Kennedy-Thorndike
  experiments''). 
  \item Occurrence of ``forbidden processes'' (such as photon decay in
  vacuum). 
  \item Dispersive processes in vacuum (such as energy dependent
  velocities or birefringence).
  \item Breakdown of discrete symmetries, principally violation of the
  CPT theorem.
\end{enumerate}

There are two main group of observations with high enough sensitivity
to probe the Quantum Gravity regime: Cosmological or astrophysical
tests of Lorentz invariance (``Threshold analysis type''), based on the
modification of the dispersion relations of particles, and Clock
comparison experiments (``Hughes-Drever type''). The sensitivity
requirements are indeed very strict; for instance, for terms linear in
the Planck length $\ell_P$, one should have an accuracy better than
\begin{equation}
  \ell_P m \sim 10^{-19} \label{equ:Accur}
\end{equation}
 for a mass of the order of the nucleon mass.

In the following we shall consider some of the most interesting tests
of Lorentz invariance. Our selection is motivated by their
effectiveness in testing Quantum Gravity induced Lorentz
violation. More complete covering of the phenomenology can be found in
references
\cite{Bluhm:2001ms,Bluhm:2001yy,Konopka:2002tt,Bluhm:2003un,%
Amelino-Camelia:2003ex,%
Bluhm:2004tm,Stecker:2004vm,Jacobson:2004rj,Jacobson:2004qt} 

\subsection{Kinematic tests of Lorentz invariance}
\label{ssec:KinTests}

These are the classical Michelson-Morley \cite{MM87} and Kennedy-Thorndike
\cite{KT32} experiments, which test  isotropy and boost
independence of the local velocity of light. Using the Robertson
metric \eqref{equ:R-Metric} it is easy to show that the light
velocity is given by
\begin{equation}
  \frac{c(\theta,V)}{c_0} = 1 + \left(\alpha + \beta -
  1\right)\frac{V^2}{c^2} +  \left(\beta - \delta -
  \frac{1}{2}\right)\frac{V^2}{c^2} \label{equ:c(theta,V)}
\end{equation}

Te last term is a local anisotropy in the velocity of light, while the
second one represents a boost dependence. Several modern versions of
both experiments have been carried in the latter years
\cite{BrilHall,HilsHall,braxmaier02,mueller03}, with strict limits on
the Mansouri-Sexl parameters
\begin{subequations}\label{equ:Vals:MSPars}
 \begin{gather}
  \left\vert \alpha + \beta - 1 \right\vert < 6.9\times10^{-7} \\
  \left\vert \delta - \beta + \frac{1}{2} \right\vert < 4.5\times10^{-9}
\end{gather} 
\end{subequations}

The result of the new version Michelson-Morley experiment
\cite{mueller03} were also interpreted also in terms of the Kosteleck\'y
$k_F$ parameter tensor. It tests only the P-even part of the tensor
and it imposes limits of
\begin{align}
  \left\vert k_{Fe^-} \right\vert &< 10^{-14} & \left\vert k_{Fo^-}
  \right\vert <  10^{-10} \label{equ:kF:Limits}
\end{align}
where the limits refer to the P-even and P-odd decomposition of $k_F$
\cite{Kostelecky:2002hh,mueller03}.

 $\alpha$ is an important parameter (or rather, the function $g_0(V)$
is) since its value is related to the dispersion relation for the
particle. Indeed, it can be shown \cite{MacArthur86} that the energy
of a moving particle is related to its velocity in the form
\begin{equation}
  E = \frac{mc^2}{g_0(V)\sqrt{1 - \frac{V^2}{c^2}}} \label{equ:E(v):R-M-S}
\end{equation}
which provides an interesting connection between the
Robertson-Mansouri-Sexl test theory and some dispersion relation
ones. But \eqref{equ:E(v):R-M-S} defines also the transverse D\"oppler
effect, from which $\alpha$ can be measured. The most accurate value
obtained is \cite{Grieser:1993qs}
\begin{equation}
  \left\vert \alpha + \frac{1}{2} \right\vert < 8\times10^{-7}
  \label{equ:alfa:R-M-S} 
\end{equation}

In spite of these impressive results, the accuracy of the above
experiments is still below the requirement for testing Quantum Gravity
at the tree level.

\subsection{``Forbidden'' processes}
\label{ssec:Forbidden}

\begin{figure}[t]
\begin{center}
\begin{minipage}[h]{80mm}
\epsfig{file=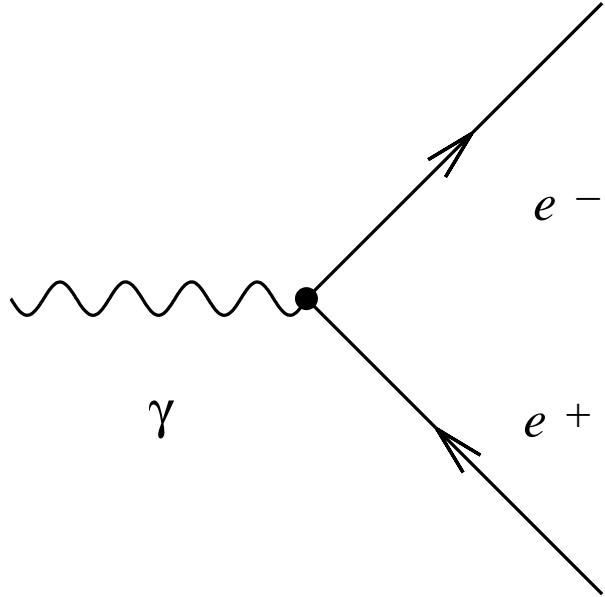, width=80mm}
\end{minipage}
\caption{The forbidden process $\gamma \to e^+ + e^-$.} \label{fig:ga2e}
\end{center}
\end{figure}

If the limiting velocity of a high energy particle is different from
the light velocity, several processes forbidden in a Lorentz invariant
context became possible \cite{Coleman:1997xq,Coleman:1998ti}. As an
example, consider the decay of a photon in an electron-positron pair,
a process which is usually forbidden by energy-momentum conservation
(Fig. \ref{fig:ga2e})
\begin{equation}
  \gamma \to e^+ + e^- \label{equ:ga2e}
\end{equation}

But if the speed of photons is greater than the limiting speed of
electrons this is no more so. Indeed, energy-momentum conservation
implies
\begin{equation}
  q = p_+ + p_- \label{equ:ga2e:EMCons}
\end{equation}

Let us examine this process in the context of the $TH\epsilon\mu$
formalism \ref{equ:S-THem:FF}. 
There is a threshold for this reaction: the photon energy should be
enough to produce an electron-positron pair at rest, so
that
\begin{equation}
  E_\gamma > \frac{2m}{\sqrt{\left(\frac{c_\gamma}{c_e}\right)^2 -1}}
  = E_0
  \label{equ:ga2e:ThrE} 
\end{equation}

Any photon with energy $E>E_0$ will rapidly decay into
electron-positron pairs so that they should never arrive to
earth. However, photons of energy $E_\gamma > \unit[20]{TeV}$ have
been observed so that an upper limit is obtained \cite{Coleman:1997xq}
\begin{equation}
  \frac{c_\gamma}{c_e} -1 < 1.5\times10^{-15}
\end{equation}
 
On the other hand, if the limiting speed is greater than the photon
 speed, the energy of charged particles is limited by vacuum Cerenkov
 process
 \begin{equation}
   p \to p + \gamma \label{equ:vCh}
 \end{equation}
whose threshold is
\begin{equation}
  E'_0 = \frac{m_p}{\sqrt{\left(1 - \frac{c_\gamma}{c_e}\right)^2 }}
\end{equation}

From the fact that the upper limit of cosmic ray spectrum is $E_u \sim
\unit[10^{20}]{eV}$ a lower bound of
\begin{equation}
  \frac{c_\gamma}{c_e} -1 > - 5\times10^{-23}
\end{equation}
was obtained \cite{Coleman:1997xq}.

Of course, the same phenomena can be analyzed in the context of any
dispersion relation model, once the effective limit- and light- speeds
can be determined. For instance, much more stringent limits have been
obtained from the $\gamma$-ray spectrum of the Crab nebula
\cite{Jacobson:2002ye}. These limits have been obtained within the
Myers-Pospelov test theory.

\subsection{Propagation phenomena: delays and polarization}
\label{ssec:TandPi}

The propagation of high energy particles, specially photons, has been
one of the most important tests of Lorentz Invariance since the
seminal paper of Amelino-Camelia \emph{et. al.}
\cite{Amelino-Camelia:1997gz}. These are based on the simple fact that
particles of different energy, with dispersion relations of the form
\eqref{def:AC:Test}, will have different velocities and so different
travel times from the same sources. 

For a pair of photons with energies $E_1, E_2$ and dispersion relation
of the form \eqref{def:AC:Test} with $n=1$ the difference in arrival
time would be
\begin{equation}
  \frac{\Delta t}{T} \simeq \eta \frac{\Delta E}{E_P}
  \label{equ:Delta-t} 
\end{equation}
where $T \simeq D/c$ is the travel time of the photon. For a photon
energy difference of $\unit[1]{GeV}$ and a travel time of
$\unit[10^{17}]{s}$ a time delay of the order of a second results,
depending on the value of $\eta$. A typical bound from $\unit{TeV}$
flares of Makarian 421 is \cite{Biller:1998hg} 
\begin{equation}
  \frac{E_P}{\eta} \ge \unit[4\times10^{16}]{GeV}
  \label{equ:Delta-t:Limit} 
\end{equation}
implying
\begin{equation}
  \eta > 300 \label{equ:Delta-t:Limit:eta}
\end{equation}

This limit is still below the accuracy requirements \eqref{equ:Accur}
for testing Quantum gravity.

On the other hand, very interesting results have been found studying
the polarization behavior of the traveling radiation. As an example,
consider the modified Maxwell equations derived from the Hamiltonian
\eqref{equ:H-GP} \cite{Gambini:1998it,Gleiser:2001rm}
\begin{subequations}\label{ecs:GP:MaxMod}
  \begin{gather}
    \dot{\bm{E}} = - \bm{\nabla\times B} +
    2\theta_{GP}\ell_P\nabla^2\bm{B}\\
    \dot{\bm{B}} = \bm{\nabla\times E} -
    2\theta_{GP}\ell_P\nabla^2\bm{B} 
  \end{gather}
\end{subequations}

These modified Maxwell equations (written in the privileged reference
system) have a solution \cite{Gambini:1998it,Gleiser:2001rm}
\begin{subequations}
  \begin{equation}
  \bm{E}_\pm = (\hat{\bm{e}}_1 \pm \hat{\bm{e}}_2)e^{i\left(\Omega_\pm
  t - \bm{k\cdot x}\right)} \label{equ:GP:PolE}
  \end{equation}
and a similar expression for $\bm{B}$. Here $\hat{\bm{e}}_{1,2}$
denote the polarization unit vectors and
\begin{equation}
  \Omega_\pm = k\left(1 \mp 2\theta_{GP} \ell_P k\right)
  \label{equ:GP:DispRel} 
\end{equation}
is the dispersion relation for photon propagation in vacuum.
\end{subequations}

The latter equation predicts birefringence of the vacuum. Indeed, for
a wave packet traveling along the $x$ axis, formed with a
superposition of plane waves with central frequency $\Omega_0$
\eqref{equ:GP:PolE}, one finds
\begin{equation}
  \bm{E} \simeq e^{i\Omega_0(t-x)} \left[\mathcal{A}(x - v_+t)
  e^{-i\theta_{GP}\ell_P\Omega_0^2z} \hat{\bm{e}_+} + \mathcal{A}(x -
  v_-t) e^{i\theta_{GP}\ell_P\Omega_0^2z} \hat{\bm{e}_-}\right]
  \label{equ:GP:Wpkt} 
\end{equation}
with $A(x)$ the initial wave packet, $\hat{\bm{e}}_\pm = \hat{\bm{e}}_1
  \pm \hat{\bm{e}}_2$ and the group velocities are given by
  \begin{equation}
    v_\pm = 1 \mp 4\theta_{GP}\ell_P k
  \end{equation}

For short enough distances, so that 
\begin{equation}
  8\theta_{GP}\ell_P\Omega_0\delta\Omega \ll 1
\end{equation}
where $\delta\Omega$ is the frequency width of the wave packet,
equation \eqref{equ:GP:Wpkt} can be recast in the form
\begin{equation}
  \bm{E} \sim B(t-x)
  \left[\cos\left(\theta_{GP}\ell_P\Omega_0^2z\right) 
  \hat{\bm{e}_1} + \sin\left(\theta_{GP}\ell_P\Omega_0^2z\right)
  \hat{\bm{e}_2} 
  \right] 
\end{equation}
showing explicitly a rotation of the polarization vector. 

From the fact that the optical polarization spectrum of many distant
galaxies is flat, the following bound was found for the Gambini-Pullin
parameter \cite{Gleiser:2001rm}
\begin{equation}
  \left\vert\theta_{GP}\right\vert < 10^{-4} \label{equ:GP:GKBound}
\end{equation}

This is a very strict bound on a parameter directly connected to
Quantum Gravity. 

Even more strict bounds were obtained for the P-odd part of the $k_F$
tensor in the Lorentz violating extension of the Standard Model
\cite{Kostelecky:2001mb,Kostelecky:2002hh}. From an analysis of a
carefully selected set of cosmologically distant sources, whose
polarization had been measured. From a careful analysis of these
sources, a limit
\begin{equation}
 \left\vert k_F^\alpha\right\vert < 10^{-31} \label{equ:KM:PolBound}
\end{equation}
was found. 

The bounds \eqref{equ:GP:GKBound} and \eqref{equ:KM:PolBound} are in a
certain sense complementary, since the  Lorentz violating extension of
the Standard Model does not include terms of dimension 5, such as
those in the Gambini-Pullin Hamiltonian. 

\subsection{Laboratory  experiments}
\label{ssec:LabExper}

There has been a great number of laboratory experiments designed to
detect tiny violations of Lorentz invariance. Among them, the
Hughes-Drever-like experiments \cite{HRBL60,Drever61} designed to
detect tiny anisotropies in the laboratory reference system, are the
most sensitive ones. 

\begin{figure}[t]
\begin{center}
\begin{minipage}[h]{80mm}
\epsfig{file=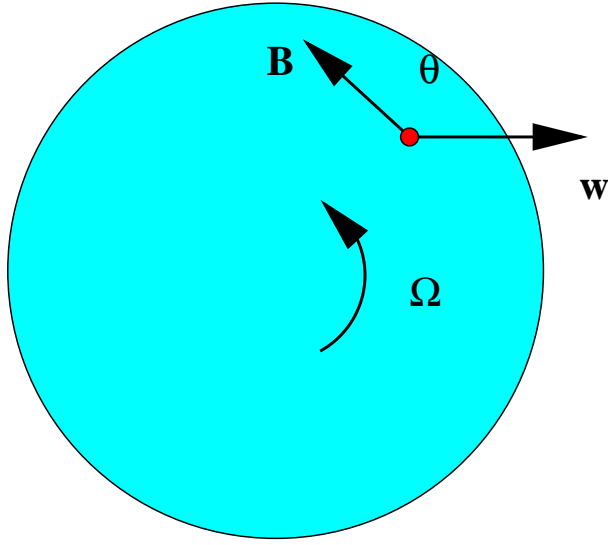, width=80mm}
\end{minipage}
\caption{Schematic view of a Hughes-Drever experiment
  \label{fig:Esq:H-D}} 
\end{center}
\end{figure}

In a Hughes-Drever experiment the spin of an atomic or nuclear system
is oriented in a magnetic field in a reference system rotating with
angular velocity $\bm{\Omega}$ and moving with velocity $\bm{w}$ with
respect of the privileged one (fig. \ref{fig:Esq:H-D}). The motion
with respect to the privileged system generates local anisotropies
that in turn generate a signal of frequency $n\Omega$. 

In the latter years many new versions of the original Hughes-Drever
experiment have been devised, with ever increasing accuracy
\cite{Prestage85,Lamoreaux89,Chupp89,Phillips01:LIV_H,Berglund95,%
Bear:2000cd,Cane:2003wp}. As an example, let us consider bounds on the
Dirac Hamiltonian parameters \eqref{equ:B6} from Loop Quantum Gravity
\cite{Sudarsky:2002ue}. 

As we have already said, the last line in \eqref{equ:B6} represent a
coupling of the spin to the velocity with respect to the privileged
frame $\bm{w}$. Keeping only the dominant terms and passing to the
usual units we find the dominant perturbation
\begin{equation}
  \delta H_S = \left(\Theta_1 + \frac{1}{2}\Theta_4\right) M\ell_P
  (2Mc^2) \left[1 + O\left(\frac{p^2}{2Mc^2}\right)\right]
  \bm{s\cdot}\frac{\bm{w}}{c} 
  \label{equ:Dirac:H-S} 
\end{equation}

The third line in \eqref{equ:B6} represents an anisotropy of the
inertial mass, to test which the Hughes-Drever experiments were
designed. With the approximation 
\begin{equation}
  Q_P \simeq - \frac{5}{3} \langle p^2\rangle \frac{Q}{R^2}
  \label{equ:Q_P} 
\end{equation}
for the momentum quadrupole moment, with $Q$ the electric quadrupole
moment and $R$ the nuclear radius, we obtain
\begin{equation}
  \delta H_Q = - \Theta_1 M\ell_P \frac{5}{3}  
  \left\langle\frac{p^2}{2M}\right\rangle  \frac{Q}{R^2}
  \left(\frac{w}{c}\right)^2 P_2(\cos\theta) \label{equ:Dirac:H-Q} 
\end{equation}
for the quadrupole moment perturbation, where $\theta$ is the angle
between the quantization axis and $\bm{w}$, $\langle p^2/2M\rangle
\sim \unit[40]{MeV}$ for the kinetic energy of the last shell of a
typical heavy nucleus. from the results of references
\cite{Chupp89,Bear:2000cd} we find \cite{Sudarsky:2002ue}
\begin{align}
  \left\vert \Theta_1 + \frac{1}{2}\Theta_4\right\vert &<
  2\times10^{-9} & \left\vert \Theta_1 \right\vert &< 3\times10^{-5}
  \label{equ::Dirac:Bounds} 
\end{align}

This set of experiments can be interpreted in different test
theories. For instance, the above results have been used to set bounds
for the Myers-Pospelov parameters \cite{Myers:2003fd}. 

Besides, the parameter $c-1$ in the $TH\epsilon\mu$ model has been
constrained from the results of references
\cite{Prestage85,Lamoreaux89,Chupp89}
\begin{equation}
  \vert c - 1 \vert < 10^{-21}
\end{equation}
the result being less sensitive than \eqref{equ::Dirac:Bounds}
because of several suppressing nuclear factors. 

Much more interesting is the possibility of constraining the mass of
the D-particle from the above results. Indeed, from equation
\eqref{eq:DiracMod} one obtains the low-energy perturbation
Hamiltonian
\begin{equation}
  H'_Q = - 4\frac{m_N}{M_D}\frac{{\bf w}\cdot {Q}_{P}\cdot {\bf
  w}}{Mc^2} \label{equ:Ellis:Pert}
\end{equation}
and comparison with \eqref{equ:B6} shows that
\begin{equation}
  M_D = \frac{4}{\Theta_1 \ell_P} > 1.2\times10^5 M_P
  \label{equ:Bound:M_D} 
\end{equation}

This is the strongest bound found so far for the mass of the
D-particle. The corresponding recoil velocity
\begin{equation}
  v < 2\times10^{-27} c
\end{equation}
is extremely small, even by the standards of everyday experience
(e.g. the speed of a crawling snail is $10^{-11}c$), that it seems
quite unlikely that it will be detected in more direct experiments,
such as time delays. 

\subsection{The GZK cutoff}
\label{ssec:GZK}

The spectrum of high energy cosmic rays should show a cutoff at
energies about $\unit[50]{EeV}$, called \emph{the GZK cutoff}
  \cite{Greisen:1966jv,Zatsepin:1966jv}.  The origin of this cutoff
comes from several processes that eat up the energy of primary cosmic
rays, such as inverse Compton effect by the Cosmic Microwave Background
photons. A modern evaluation of the ultrahigh energy spectrum of cosmic
rays can be found in \cite{Scully:2000dr} and a complete review of the
problem in \cite{Stecker:2002fh}. 

In spite of the theoretical prediction, quite a few cosmic rays of
energies greater than the GZK cutoff have been found in several
cosmic ray observatories, mainly AGASA
\cite{Takeda:1998ps,Takeda:1999sg}. This is usually called \emph{the
  GZK anomaly}.

\begin{figure}[t]
\begin{center}
\begin{minipage}[h]{80mm}
\epsfig{file=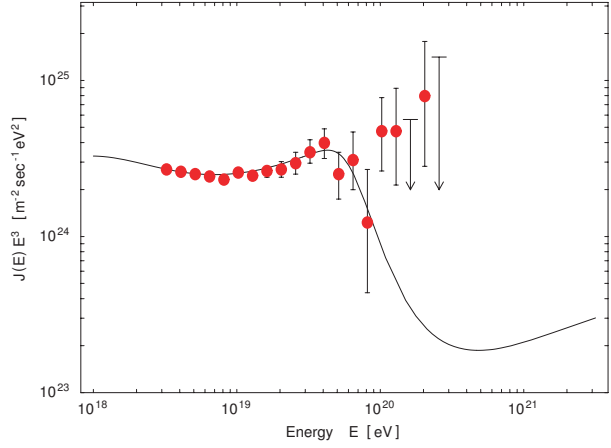, width=80mm}
\end{minipage}
\caption{The GZK cutoff and the AGASA High energy data.}
\label{fig:Alfaro1} 
\end{center}
\end{figure}

Alfaro and Palma \cite{Alfaro:2001gk,Alfaro:2002ya} have shown that an
explanation of the GZK anomaly can be found within the Loop Quantum
gravity model, and besides, that this can be used to test quadratic
terms in the Planck scale. 

To see how this can be done, observe that the dispersion relations for
particles in the  Loop Quantum Gravity of section \ref{ssec:LIV-LQG}
take the form
\begin{subequations}\label{ecs:LQG:DispRels}
  \begin{gather}
  E_\pm^2 \simeq \left[1 + 2
  \kappa_\alpha\left(\frac{\ell_P}{\CL}\right)^2 \right] p^2 +
  \kappa_\eta \ell_P^2 p^4 \pm \kappa_\lambda\frac{\ell_P}{2\CL^2}p +
  m^2 \\
  \omega_\pm \simeq k\left[1 +
  \kappa_\gamma\left(\frac{\ell_P}{\CL}\right)^{2+2\Upsilon} -
  \theta_3\left(\ell_Pk\right)^2 \pm \theta_{GP} \ell_Pk\right]
\end{gather}
\end{subequations}
for massive fermions and photons respectively. In these equations the
$\kappa_i$ and $\theta$ parameters are numbers of order unity. The
explicit dependence with the \CL\ and $\Upsilon$ parameters has been
left explicit.

With these equations, a careful analysis of the main reactions in the
presence of Lorentz violations was carried with the result that the
GZK anomaly could be explained. A fit to the data is possible assuming
that \CL\ is an universal constant with a value
\begin{equation}
  \unit[2.6\times10^{-18}]{eV^{-1}} \lesssim \CL \lesssim
  \unit[1.6\times10^{-17}]{eV^{-1}} 
\end{equation}

\begin{figure}[t]
\begin{center}
\begin{minipage}[h]{80mm}
\epsfig{file=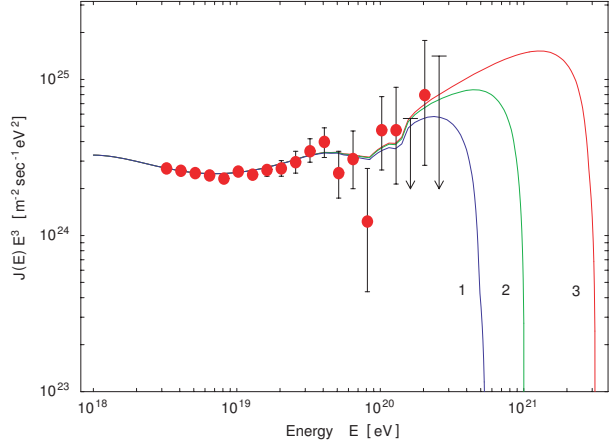, width=80mm}
\end{minipage}
\caption{Lorentz Noninvariant fit of the GZK anomaly} \label{fig:Alfaro4}
\end{center}
\end{figure}

The good fit shown in Figure \ref{fig:Alfaro4} shows the importance of
threshold analysis for the test of even quadratic order in the Planck
scale. 

\subsection{Other tests of  Lorentz invariance}
\label{ssec:OtherTests}

There are many other tests of Lorentz invariance, but few of them
satisfy our accuracy requirement \eqref{equ:Accur}. The most important
of them are the laboratory test for CPT symmetry
\cite{Dehmelt:1999jh,Mittleman:1999it,Gabrielse:1999kc}.  

These experiments test the CPT symmetry of an electron (proton)
in a Penning trap. Although potentially they should be very sensitive,
they have not yet our required accuracy \eqref{equ:Accur}. Their
impressive result
\begin{equation}
  \frac{\vert\Delta a\vert \bar\omega_c}{2m_ec^2} =
  (3\pm12)\times10^{-22}  \label{equ:CPT-Test}
\end{equation}
translates into the weak constraint \cite{Bluhm:2001yy}
\begin{equation}
  \left\vert\Theta_2 + \frac{1}{2}\Theta_4\right\vert \lesssim 1
  \label{equ:CPT-Bound} 
\end{equation}

Another important group of experiments are made with spin-polarized
matter \cite{Heckel:1999sy,Bluhm:2001yy}. In these experiments a
torsion pendulum with a total spin $S\sim 10^{22}$ can be used to
extract a signal coupling spin and velocity with respect to a
privileged frame. Their result translates into the mild bound
\begin{equation}
  \left\vert\Theta_2 + \frac{1}{2}\Theta_4\right\vert \lesssim 0.002
  \label{equ:Spin-Bound} 
\end{equation}

Although the increasing accuracy of these experiments will improve
very much the above bound, they are still far from the results of
Hughes-Drever experiments. Their main interest is that the parameters
$\Theta_i$ are estimated from electron instead of proton, yielding
independent tests of Lorentz invariance.

\section{The influence of radiative corrections}
\label{sec:RadCorr}

\begin{figure}[t]
\begin{center}
\begin{minipage}[h]{80mm}
\epsfig{file=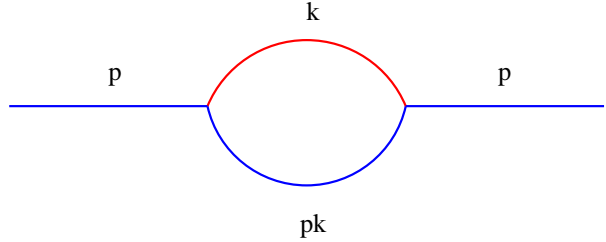, width=80mm}
\end{minipage}
\caption{A typical self-energy diagram} \label{fig:SelfEn}
\end{center}
\end{figure}

Radiative corrections, whether  at one-loop or higher order, open the
possibility of testing energies as high as Planck's in nowadays
existing  laboratory experiments. Indeed, consider the diagram in
Figure \ref{fig:SelfEn}: the loop variable $k$ runs over all energies,
probing thus near Planck scale.

Although quite a lot of work has been done on radiative corrections in
Lorentz non-invariant theories
\cite{Kostelecky:2001jc,Carroll:2001ws,Burgess:2002tb,Perez:2003un}
only quite recently it was shown that they impose very stringent
constraints on Lorentz violation phenomena \cite{Collins:2004bp}. In
the context of renormalization theory, this originates a new fine
tuning problem.

The main idea is that Planck scale physics should cutoff the range of
the loop variable at a given scale $\Lambda \sim E_P$. If this cutoff
is anisotropic, this anisotropy will be ``dredged up'' to a low energy
scale. This would include corrections to the dispersion relations of
the order of the coupling constants of the standard model, rather than
$O(\ell_P p)$.

In the following sections, we shall discuss first a couple of examples
and then discuss the general case.

\subsection{A couple of examples}
\label{ssec:Examples}

Let us consider first the electron self-energy, introducing a cutoff
function $C(k^2)$ to model Planck scale effects
\begin{equation}
  \Sigma^{(2)}(p,\Lambda,\xi) = \frac{e^2}{4\pi i} \int d^4k
  \gamma^\mu \frac{\gamma\cdot(p-k)+m}{(p-k)^2-m^2} \gamma_\mu
  \frac{1}{k^2+i\epsilon} C\left(k^2 + \xi(W\cdot k)^2\right)
  \label{equ:Sigma-e} 
\end{equation}
where $\xi$ parametrizes an anisotropy in the laboratory
system. Consider the particular case
\begin{equation}
  C(k^2) = \frac{\Lambda^2}{\Lambda^2 - k^2}
\end{equation}
with $\Lambda \sim E_P$ and $\xi \lesssim 1$. Then we can expand the self
energy in powers of $\xi$ and keep the lowest order term
\begin{equation}
  \Sigma^{(2)}(p,\Lambda,\xi) = \Sigma^{(2)}(p,\Lambda,0) +
  \delta\Sigma \label{def:delta-Sigma}
\end{equation}
with
\begin{equation}
  \delta\Sigma(p,\Lambda,\xi) = -i\alpha \int d^4k \gamma^\mu
  \frac{\gamma\cdot(p-k)+m}{(p-k)^2-m^2} \gamma_\mu
  \frac{1}{k^2+i\epsilon} \frac{\Lambda^2}{\Lambda^2 - k^2}
  \frac{\xi(W\cdot k)^2}{\Lambda^2 - k^2} \label{equ:Int:delta-Sigma} 
\end{equation}

After the usual procedure of combining denominators and shifting the
loop variable one is lead to the on-shell expression
\begin{equation}
  \delta\Sigma(\not{p}=m,\Lambda,\xi) = 12\alpha\xi \int_0^1zdz
  \int_0^1y^2dy \int d^4q \frac{P(q)\Lambda^2}{\left[q^2 - m^2(1-y)^2
  - \Lambda^2xy\right]} 
\end{equation}
with
\begin{equation}
  P(q) = \frac{1}{4}\left[(4+2y)mW^2 - 4(\gamma\cdot W)(W\cdot
  p)(1-y)\right]q^2 + O(q^0) 
\end{equation}

To compute this integral one usually performs a Wick rotation, but it
is not clear that this can be done in the highly fractal structure of
spacetime at the Planck scale. Indeed, one expects that the physical
cutoff induced by discreteness will be in real spacetime. But this can
be done in this particular model with the result
\begin{equation}
  \delta\Sigma(\not{p}=m,\Lambda,\xi) = \pi^2\alpha\xi \int_0^1zdz
  \int_0^1y^2dy \tilde{P}(y)
  \frac{\Lambda^2}{m^2(1-y)^2 + \Lambda^2xy}
\end{equation}
with
\begin{equation}
  \tilde{P}(y) = [(4+2y)mW^2 - 2(\gamma\cdot W)(W\cdot  p)(1-y)]
\end{equation}

The integral is finite in the limit $\Lambda \to \infty$ and we find
thus a correction term in the nonrelativistic limit 
\begin{equation}
  \delta\Sigma \sim \xi\alpha m (\bm{s\cdot w})
\end{equation}
that contradicts experiment unless $\xi \ll 10^{-28}$. 

As a second example, let us consider a non-commutative version of QED,
with
\begin{equation}
  \left[x^\mu,x^\nu\right] = i \theta^{\mu\nu}
\end{equation}
with $\theta^{\mu\nu}$ a $c$-number.

This case has been analyzed in reference \cite{Anisimov:2001zc} with
the result
\begin{equation}
  \delta\CL_{\rm ef} = 
  \begin{cases}
    \frac{3}{4} m\Lambda^2 \alpha^2 \bar{\psi}
    \theta^{\mu\nu}\sigma_{\mu\nu} \psi, & \theta\Lambda^2 \ll 1\\
    2\pi m\alpha^2
    \bar{\psi}
    \frac{\theta^{\mu\nu}\sigma_{\mu\nu}}{\frac{1}{2}\operatorname{Tr}\theta^2}
    \psi, & \theta\Lambda^2 \gg 1
  \end{cases}
\end{equation}
where $\theta$ is a typical scale of non-commutativity, such as $\theta
= \sqrt{\theta^{\mu\nu}\theta_{\mu\nu}}$. 

 The second result predicts an anisotropy of order $\alpha^2$,
independent of both the scale of non-commutativity and cutoff and thus
is in strong disagreement with experiment, while the first can only be
consistent in the strange situation where the cutoff $\Lambda$ is much
smaller than the interesting non-commutative scale. This means, of
course that the non-commutative field theory ceases to be valid at
energy scales much smaller than the non-commutativity scale.

\begin{figure}[t]
\begin{center}
\begin{minipage}[h]{80mm}
\epsfig{file=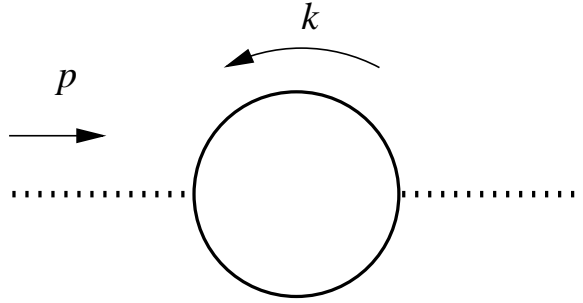, width=80mm}
\end{minipage}
\caption{Lowest order self-energy graph in a Yukawa theory.}
\label{fig:ScSelf}
\end{center}
\end{figure}

For our last example, consider the second order self energy of a
scalar particle in a Yukawa theory $\Pi^{(2)}(p)$ (Figure
\ref{fig:ScSelf}). A measure of Lorentz violating in this graph is
given by the quantity \cite{Collins:2004bp}
\begin{equation}
  \xi = \left.\frac{\partial^2 \Pi^{(2)}(p)}{\partial p_0^2} +
  \frac{\partial^2 \Pi^{(2)}(p)}{\partial p_1^2}\right\vert_{p=0}
  \label{def:our:xi}
\end{equation}

Without Planck scale modifications one would obtain
\begin{equation}
  \xi = -\frac{ig^2}{\pi^4} \int d^4k \frac{\left(k_0^2+k_1^2\right)
  \left(k^2+3M^2\right) }{\left(k^2 - M^2 + i\epsilon\right)^4}
  \label{equ:xi-Liv} 
\end{equation}

This integral could be shown to be zero by a Wick rotation, after a
suitable regularization to avoid the logarithmic divergence. However,
if Planck scale physics introduces a cutoff, things are different. For
instance,  assume that the fermion propagator has a cutoff in the form
\begin{align}
  S'(k) &= f(\bm{k})\frac{\gamma\cdot k + M}{k^2 - M^2} & f(0) &=1 &
  f(\infty) &= 0
\end{align}
then the integral can be computed and the corresponding Lorentz
violation is
\begin{equation}
  \xi = \frac{g^2}{6\pi^2}\left[1 + \int_0^\infty dx x f'(x)^2\right]
  \label{equ:xi:value} 
\end{equation}

This shows again that a Lorentz violation at the Planck sale is
translated into a Lorentz violating term at low energy, of the order
of the coupling constant squared. 

\subsection{A general argument}
\label{ssec:GenRen}

Let us now show that the above examples are really parts of a general
property of radiative corrections.
To fix our ideas, consider the self energy of a scalar particle in
a Yukawa theory. The dispersion relation for the particle is obtained
by solving
\begin{equation}
  E^2 - \bm{p}^2 - m^2 - \Pi(E,\bm{p}) = 0 \label{equ:DispRelY}
\end{equation}
with $\Pi$ the sum of all self-energy graphs, to which we have added
any small Lorentz-violating corrections that stem from tree-level theory. 

Now we reason as follows: without a cutoff the graphs have divergences
from large momenta. In the Lagrangian defining the theory the
divergences correspond to terms of dimension 4 (or less) that obey the
symmetries of the microscopic theory. In a Lorentz invariant case
these divergences are removed by renormalization of the parameters of
the theory. But Planck scale physics can both cutoff the divergences
and modify the expressions that define the graph in the neighborhood of
the Planck energy. The same power counting that determines the
divergences also determines the natural size of the contributions from
Planck scale momenta: the dominant contributions correspond to
operators of dimension 4 (or less) in the Lagrangian. If the
microscopic theory violates Lorentz invariance at the Planck scale,
then generically we get Lorentz violating terms at low energy, without
any suppression by powers of $E/E_P$.  

Of course, the above Lorentz violating terms can be removed by
explicitly including Lorentz violating counterterms of dimension 4 in
the Lagrangian which are fine-tuned to give the observed low-energy
Lorentz invariance. Such fine-tuning is unacceptable
\cite{Weinberg73,Susskind:1978ms,Weinberg:1988cp} in a fundamental
theory.

In the models considered in Section \ref{sec:TestTh} and
\ref{sec:LIVTest} Lorentz violating corrections to the dispersion
relations, suppressed by some power of  $E/E_P$, were found by
considering the propagation of free particles in a granular spacetime
background. The above reasoning shows that there are effects that are
only suppressed by two powers of the standard model coupling
constants. These effects change the effective value of $c$ in the
usual dispersion relation
\begin{equation}
  E^2 = c^2p^2 + m^2c^4 \label{equ:DispRel:Usual}
\end{equation}

The well-known structure of the standard model shows that
\eqref{equ:DispRel:Usual} predicts differences up to 10\% between
different particles, in violent disagreement with experiment that
shows that the fractional differences in $c$ is below $10^{-20}$.

The above result is a consequence of well-known properties of quantum
field theories, of which the standard model is an example. Indeed the
actual technical result is related to more or less explicit statements
that can be found in several references
\cite{Kostelecky:1994rn,Colladay:1996iz,Coleman:1998ti,Colladay:1998fq,%
Burgess:2002tb,Myers:2003fd} although never fully recognized except n
the recent \cite{Myers:2004ge}. For a criticism of these results see
\cite{Perez:2003un,Collins:2004bp}.

\section{Conclusion}
\label{sec:Concl}

In spite of the difficult nature of the research, a group of delicate
experiments has been devised to test Lorentz invariance violations
induced by Quantum Gravity. Up to now, no clear signal of Lorentz
violation has been found (See, however, Section \ref{ssec:GZK}).

Tree-level tests of Quantum Gravity modifications include attempts to
measure the parameters of the effective Lagrangian's describing these
modifications.  Several of the above mentioned experiments are
accurate enough to measure the parameters of corrected dispersion
relations with high precision. Even better results can be found from
Hughes-Drever experiments. Indeed, the bounds on the parameters are so
small that they call into question the full scheme of Lorentz
invariance violations induced by Quantum Gravity, at least with
respect to terms linear in the Planck length 
\cite{Sudarsky:2002ue,Sudarsky:2002zy}. 

Even more strict results stem from the analysis of radiative
corrections. We have found that the loop variables can dredge up 
Lorentz violations at the Planck scale up to the low-energy
scale. This would violently contradict experiment, unless extremely
fine tuning counterterms are introduced in the Lorentz violating
Lagrangian. This shows that Lorentz invariance should be added to the
well-known list of fine-tuning problems; namely, the cosmological
constant, the Higgs bare mass and mass hierarchies. 

The implications of the above argument for both experiment and  theory
are quite profound. First, the present unsuccessful searches for
Lorentz violations suffice by many orders of magnitude. Of course,
since it is correct science to question and test accepted principles,
tests of Lorentz invariance are worthwhile. However, many searches
that were started by estimates of specific orders of magnitude may be
misled and should be revised.

As to the theory, the critical task concerns any proposal in which
Lorentz symmetry is substantially broken at the Planck scale: to find
and implement a mechanism to give automatic Lorentz invariance at low
energy despite a violation at Planck scale. We assume here that the
treatment involves real time, not an analytic continuation to
imaginary time, as is common in quantum field theory in flat spacetime
(See, however, Section \ref{ssec:Examples}). One mechanism is to have
a custodial symmetry that is sufficient to prohibit Lorentz-violating
symmetry without being itself the full Lorentz group. Such a symmetry
does not appear to be known and the Coleman-Mandula theorem
\cite{Coleman:1967ad} suggests it does not exists.

It should be noted that there is not necessarily a conflict between
discreteness and the absence of a preferred frame. For instance in
reference \cite{Dowker:2003hb} a Lorentz invariant macroscopic space
is constructed by the use of a random causal set of points. On the
other hand, the authors of \cite{Rovelli:2002vp} argue that the
existence of a minimum length does not imply local Lorentz invariance
violation, anymore than the discreteness of angular momentum
eigenvalues signal a violation of rotational invariance. 

An optimistic point of view should be stressed: a branch of theoretical
physics long considered to suffer from detachment from experimental
guidance is now in the opposite situation. The subtle interplay of
cosmology, atomic and nuclear physics is shedding light on such a
recondite subject as quantum gravity. 

Our results show that Lorentz invariance still imposes stringent
requirements on the mathematical theories hat attempt to describe
experimental results.






\begin{center}
{\bf Acknowledgements}
\end{center}

The author would like to thank NovaScience for its invitation to write
this paper. This work has been partially sponsored by the Project
42026-F of CONACYT, M\'exico and Grant 11/G071 Of Universidad de La
Plata, Argentina. 

\bibliography{LIV}

\begin{thebibliography}{100}
\providecommand{\enquote}[1]{``#1''}
\expandafter\ifx\csname url\endcsname\relax
  \def\url#1{\texttt{#1}}\fi
\expandafter\ifx\csname urlprefix\endcsname\relax\def\urlprefix{URL }\fi

\bibitem{Kostelecky:1988zi}
Kostelecky, V.~A., and Samuel, S., \emph{Phys. Rev.}, \textbf{D39}, 683 (1989).

\bibitem{Kostelecky:1989jp}
Kostelecky, V.~A., and Samuel, S., \emph{Phys. Rev. Lett.}, \textbf{63}, 224
  (1989).

\bibitem{Kostelecky:1989jw}
Kostelecky, V.~A., and Samuel, S., \emph{Phys. Rev.}, \textbf{D40}, 1886--1903
  (1989).

\bibitem{Kostelecky:1991ak}
Kostelecky, V.~A., and Potting, R., \emph{Nucl. Phys.}, \textbf{B359}, 545--570
  (1991).

\bibitem{Kostelecky:1995qk}
Kostelecky, V.~A., and Potting, R., \emph{Phys. Lett.}, \textbf{B381}, 89--96
  (1996).

\bibitem{Colladay:1996iz}
Colladay, D., and Kostelecky, V.~A., \emph{Phys. Rev.}, \textbf{D55},
  6760--6774 (1997).

\bibitem{Colladay:1998fq}
Colladay, D., and Kostelecky, V.~A., \emph{Phys. Rev.}, \textbf{D58}, 116002
  (1998).

\bibitem{Amelino-Camelia:1997gz}
Amelino-Camelia, G., Ellis, J.~R., Mavromatos, N.~E., Nanopoulos, D.~V., and
  Sarkar, S., \emph{Nature}, \textbf{393}, 763--765 (1998).

\bibitem{Amelino-Camelia:2001dy}
Amelino-Camelia, G., \emph{Nature}, \textbf{410}, 1065--1069 (2001).

\bibitem{Ahluwalia:1999aj}
Ahluwalia, D.~V., \emph{Nature}, \textbf{398}, 199 (1999).

\bibitem{Amelino-Camelia:2004hm}
Amelino-Camelia, G., Introduction to quantum-gravity phenomenology (2004),
  gr-qc/0412136.

\bibitem{sorkin91}
Sorkin, R.~D., \enquote{Space-time and causal sets,} in \emph{Procedings of
  Silarg VII}, edited by J.~C. d'Olivo et~al., World Scientific, Singapore,
  1991, p. 150.

\bibitem{Gambini:1998it}
Gambini, R., and Pullin, J., \emph{Phys. Rev.}, \textbf{D59}, 124021 (1999).

\bibitem{Alfaro:1999wd}
Alfaro, J., Morales-Tecotl, H.~A., and Urrutia, L.~F., \emph{Phys. Rev. Lett.},
  \textbf{84}, 2318--2321 (2000).

\bibitem{Alfaro:2001rb}
Alfaro, J., Morales-Tecotl, H.~A., and Urrutia, L.~F., \emph{Phys. Rev.},
  \textbf{D65}, 103509 (2002).

\bibitem{Alfaro:2002xz}
Alfaro, J., Morales-Tecotl, H.~A., and Urrutia, L.~F., \emph{Phys. Rev.},
  \textbf{D66}, 124006 (2002).

\bibitem{Urrutia:2004pu}
Urrutia, L.~F., Flat space modified particle dynamics induced by loop quantum
  gravity (2004), hep-ph/0402271.

\bibitem{Ellis:1999uh}
Ellis, J.~R., Mavromatos, N.~E., and Nanopoulos, D.~V., \emph{Gen. Rel. Grav.},
  \textbf{32}, 127--144 (2000).

\bibitem{Ellis:1999sf}
Ellis, J.~R., Mavromatos, N.~E., Nanopoulos, D.~V., and Volkov, G., \emph{Gen.
  Rel. Grav.}, \textbf{32}, 1777--1798 (2000).

\bibitem{Ellis:1999sd}
Ellis, J.~R., Farakos, K., Mavromatos, N.~E., Mitsou, V.~A., and Nanopoulos,
  D.~V., \emph{Astrophys. J.}, \textbf{535}, 139--151 (2000).

\bibitem{Ellis:2000dy}
Ellis, J.~R., Mavromatos, N.~E., and Nanopoulos, D.~V., \emph{Phys. Rev.},
  \textbf{D63}, 024024 (2001).

\bibitem{Seiberg:1999vs}
Seiberg, N., and Witten, E., \emph{JHEP}, \textbf{09}, 032 (1999).

\bibitem{Douglas:2001ba}
Douglas, M.~R., and Nekrasov, N.~A., \emph{Rev. Mod. Phys.}, \textbf{73},
  977--1029 (2001).

\bibitem{grandi02:PhDTh}
Grandi, N.~E., \emph{Teor{\'\i}as de gauge no conmutativas: {C}hern-{S}immons y
  {B}orn-{I}nfeld}, Ph.D. thesis, U. N. L. P. (2002), (In spanish).

\bibitem{Sudarsky:2002ue}
Sudarsky, D., Urrutia, L., and Vucetich, H., \emph{Phys. Rev. Lett.},
  \textbf{89}, 231301 (2002).

\bibitem{Sudarsky:2002zy}
Sudarsky, D., Urrutia, L., and Vucetich, H., \emph{Phys. Rev.}, \textbf{D68},
  024010 (2003).

\bibitem{COBE}
Lineweaver, C.~H., The {CMB} dipole: The most recent measurement and some
  history (1996), \texttt{astro-ph/9609034}.

\bibitem{Streater68}
Streater, R.~F., and Wightman, A.~S., \emph{{PCT}, Spin {\&} Statistics, and
  all that}, Benjamin, New York, 1968.

\bibitem{Weinberg95}
Weinberg, S., \emph{The quantum theory of fields: Foundations}, vol.~I,
  Cambridge U. P., 1995.

\bibitem{Greenberg:2002uu}
Greenberg, O.~W., \emph{Phys. Rev. Lett.}, \textbf{89}, 231602 (2002).

\bibitem{Robertson49}
Roberson, H.~P., \emph{Rev. Mod. Phys.}, \textbf{21}, 378 (1949).

\bibitem{Mansouri77a}
Mansouri, R., and Sexl, R.~U., \emph{Gen. Rel. and Grav.}, \textbf{8}, 497
  (1977).

\bibitem{Mansouri77b}
Mansouri, R., and Sexl, R.~U., \emph{Gen. Rel. and Grav.}, \textbf{8}, 515
  (1977).

\bibitem{Mansouri77c}
Mansouri, R., and Sexl, R.~U., \emph{Gen. Rel. and Grav.}, \textbf{8}, 809
  (1977).

\bibitem{MacArthur86}
MacArthur, D.~W., \emph{Phys. Rev. A}, \textbf{33}, 1 (1986).

\bibitem{MM87}
Michelson, A.~A., and Morley, E.~W., \emph{Am. J. Sci.}, \textbf{34}, 333
  (1887).

\bibitem{KT32}
Kennedy, R.~J., and Thorndike, E.~M., \emph{Phys. Rev.}, \textbf{42}, 400
  (1932).

\bibitem{YS38}
Yves, H.~E., and Stillwell, G.~R., \emph{J. Opt. Soc. Am.}, \textbf{28}, 215
  (1938).

\bibitem{Amelino-Camelia:2004ht}
Amelino-Camelia, G., \emph{New J. Phys.}, \textbf{6}, 188 (2004).

\bibitem{Konopka:2002tt}
Konopka, T.~J., and Major, S.~A., \emph{New J. Phys.}, \textbf{4}, 57 (2002).

\bibitem{Alfaro:2002ya}
Alfaro, J., and Palma, G., \emph{Phys. Rev.}, \textbf{D67}, 083003 (2003).

\bibitem{Lehnert:2003ue}
Lehnert, R., \emph{Phys. Rev.}, \textbf{D68}, 085003 (2003).

\bibitem{LightLee73}
Lightman, A.~P., and Lee, D.~L., \emph{Phys. Rev. D}, \textbf{8}, 364 (1973).

\bibitem{GMSM}
Horvath, J., Logi{\'u}dice, E.~A., Riveros, C., and Vucetich, H., \emph{Phys.
  Rev. D}, \textbf{38}, 1754 (1988).

\bibitem{Kostelecky:1999rh}
Kostelecky, V.~A., {L}orentz- and {CPT}-violating extension of the standard
  model (1999), \texttt{hep-ph/9912528}.

\bibitem{Kostelecky:2003fs}
Kostelecky, V.~A., \emph{Phys. Rev.}, \textbf{D69}, 105009 (2004).

\bibitem{Kostelecky:2001mb}
Kostelecky, V.~A., and Mewes, M., \emph{Phys. Rev. Lett.}, \textbf{87}, 251304
  (2001).

\bibitem{Kostelecky:2002hh}
Kostelecky, V.~A., and Mewes, M., \emph{Phys. Rev.}, \textbf{D66}, 056005
  (2002).

\bibitem{mueller03}
M{\"u}ller, H., Herrmann, S., Braxmaier, C., Schiller, S., and Peters, A.,
  \emph{Phys. Rev. Lett.}, \textbf{91}, 020401 (2003).

\bibitem{Kostelecky:1999zh}
Kostelecky, V.~A., and Lane, C.~D., \emph{J. Math. Phys.}, \textbf{40},
  6245--6253 (1999).

\bibitem{Kostelecky:2000mm}
Kostelecky, V.~A., and Lehnert, R., \emph{Phys. Rev.}, \textbf{D63}, 065008
  (2001).

\bibitem{Kostelecky:2001jc}
Kostelecky, V.~A., Lane, C.~D., and Pickering, A. G.~M., \emph{Phys. Rev.},
  \textbf{D65}, 056006 (2002).

\bibitem{Bailey:2004na}
Bailey, Q.~G., and Kostelecky, V.~A., \emph{Phys. Rev.}, \textbf{D70}, 076006
  (2004).

\bibitem{Montemayor:2004uy}
Montemayor, R., and Urrutia, L.~F., Radiation in {L}orentz violating
  electrodynamics (2004), \texttt{hep-ph/0412023}.

\bibitem{Bluhm:1997ci}
Bluhm, R., Kostelecky, V.~A., and Russell, N., \emph{Phys. Rev. Lett.},
  \textbf{79}, 1432--1435 (1997).

\bibitem{Bluhm:1998rk}
Bluhm, R., Kostelecky, V.~A., and Russell, N., \emph{Phys. Rev. Lett.},
  \textbf{82}, 2254--2257 (1999).

\bibitem{Bluhm:1999ev}
Bluhm, R., and Kostelecky, V.~A., \emph{Phys. Rev. Lett.}, \textbf{84},
  1381--1384 (2000).

\bibitem{Bluhm:2001rw}
Bluhm, R., Kostelecky, V.~A., Lane, C.~D., and Russell, N., \emph{Phys. Rev.
  Lett.}, \textbf{88}, 090801 (2002).

\bibitem{Bluhm:2001ms}
Bluhm, R., Electromagnetic tests of {L}orentz and {CPT} symmetry (2001),
  \texttt{hep-ph/0112318}.

\bibitem{Bluhm:2003un}
Bluhm, R., Kostelecky, V.~A., Lane, C.~D., and Russell, N., \emph{Phys. Rev.},
  \textbf{D68}, 125008 (2003).

\bibitem{Bluhm:2003ne}
Bluhm, R., \emph{Nucl. Instrum. Meth.}, \textbf{B221}, 6--11 (2004).

\bibitem{Bluhm:2004tm}
Bluhm, R., {QED} tests of {L}orentz symmetry (2004), \texttt{hep-ph/0411149}.

\bibitem{Myers:2003fd}
Myers, R.~C., and Pospelov, M., \emph{Phys. Rev. Lett.}, \textbf{90}, 211601
  (2003).

\bibitem{Myers:2004ge}
Myers, R.~C., and Pospelov, M., Experimental challenges for quantum gravity
  (2004), gr-qc/0402028.

\bibitem{Perez:2003un}
Perez, A., and Sudarsky, D., \emph{Phys. Rev. Lett}, \textbf{91}, 179101
  (2003).

\bibitem{Collins:2004bp}
Collins, J., Perez, A., Sudarsky, D., Urrutia, L., and Vucetich, H.,
  \emph{Phys. Rev. Lett.}, \textbf{93}, 191301 (2004).

\bibitem{Magueijo:2001cr}
Magueijo, J., and Smolin, L., \emph{Phys. Rev. Lett.}, \textbf{88}, 190403
  (2002).

\bibitem{Amelino-Camelia:2002wr}
Amelino-Camelia, G., \emph{Nature}, \textbf{418}, 34--35 (2002).

\bibitem{Amelino-Camelia:2003ex}
Amelino-Camelia, G., Kowalski-Glikman, J., Mandanici, G., and Procaccini, A.,
  Phenomenology of doubly special relativity (2003), \texttt{gr-qc/0312124}.

\bibitem{Carmona:2002iv}
Carmona, J.~M., Cortes, J.~L., Gamboa, J., and Mendez, F., \emph{Phys. Lett.},
  \textbf{B565}, 222--228 (2003).

\bibitem{GrootNibbelink:2004za}
Groot~Nibbelink, S., and Pospelov, M., Lorentz violation in supersymmetric
  field theories (2004), \texttt{hep-ph/0404271}.

\bibitem{Thiemann:2002nj}
Thiemann, T., \emph{Lect. Notes Phys.}, \textbf{631}, 41--135 (2003).

\bibitem{Perez:2004hj}
Perez, A., Introduction to loop quantum gravity and spin foams (2004),
  \texttt{gr-qc/0409061}.

\bibitem{Smolin:2004sx}
Smolin, L., An invitation to loop quantum gravity (2004),
  \texttt{hep-th/0408048}.

\bibitem{Nicolai:2005mc}
Nicolai, H., Peeters, K., and Zamaklar, M., Loop quantum gravity: an outside
  view (2005), \texttt{hep-th/0501114}.

\bibitem{Ashtekar:1986yd}
Ashtekar, A., \emph{Phys. Rev. Lett.}, \textbf{57}, 2244--2247 (1986).

\bibitem{Ashtekar:1987gu}
Ashtekar, A., \emph{Phys. Rev.}, \textbf{D36}, 1587--1602 (1987).

\bibitem{Gambini:1980wm}
Gambini, R., and Trias, A., \emph{Phys. Rev.}, \textbf{D22}, 1380 (1980).

\bibitem{Gambini:1980yz}
Gambini, R., and Trias, A., \emph{Phys. Rev.}, \textbf{D23}, 553 (1981).

\bibitem{diBartolo:1983pt}
di~Bartolo, C., Nori, F., Gambini, R., and Trias, A., \emph{Nuovo Cim. Lett.},
  \textbf{38}, 497 (1983).

\bibitem{Bojowald:2003uh}
Bojowald, M., and Morales-Tecotl, H.~A., \emph{Lect. Notes Phys.},
  \textbf{646}, 421--462 (2004).

\bibitem{Bojowald:2004ax}
Bojowald, M., \emph{Pramana}, \textbf{63}, 765--776 (2004).

\bibitem{Ellis:2002bu}
Ellis, J.~R., Gravanis, E., Mavromatos, N.~E., and Nanopoulos, D.~V., Impact of
  low-energy constraints on lorentz violation (2002), \texttt{gr-qc/0209108}.

\bibitem{Bluhm:2001yy}
Bluhm, R., Probing the planck scale in low-energy atomic physics (2001),
  \texttt{hep-ph/0111323}.

\bibitem{Stecker:2004vm}
Stecker, F.~W., High energy astrophysics tests of {L}orentz invariance
  violation (2004), \texttt{astro-ph/0409731}.

\bibitem{Jacobson:2004rj}
Jacobson, T., Liberati, S., and Mattingly, D., Astrophysical bounds on {P}lanck
  suppressed {L}orentz violation (2004), \texttt{hep-ph/0407370}.

\bibitem{Jacobson:2004qt}
Jacobson, T., Liberati, S., and Mattingly, D., Quantum gravity phenomenology
  and {L}orentz violation (2004), \texttt{gr-qc/0404067}.

\bibitem{BrilHall}
Brillet, A., and Hall, J.~L., \emph{Phys. Rev. Lett.}, \textbf{42}, 549 (1979).

\bibitem{HilsHall}
Hils, D., and Hall, J.~L., \emph{Phys. Rev. Lett.}, \textbf{64}, 1697 (1990).

\bibitem{braxmaier02}
Braxmaier, C., M{\"u}ller, H., Pradl, O., Mlynck, J., and Peters, A.,
  \emph{Phys. Rev. Lett.}, \textbf{88}, 010401 (2002).

\bibitem{Grieser:1993qs}
Grieser, R., et~al., \emph{Appl. Phys.}, \textbf{B59}, 127--133 (1994).

\bibitem{Coleman:1997xq}
Coleman, S.~R., and Glashow, S.~L., \emph{Phys. Lett.}, \textbf{B405}, 249--252
  (1997).

\bibitem{Coleman:1998ti}
Coleman, S.~R., and Glashow, S.~L., \emph{Phys. Rev.}, \textbf{D59}, 116008
  (1999).

\bibitem{Jacobson:2002ye}
Jacobson, T., Liberati, S., and Mattingly, D., \emph{Nature}, \textbf{424},
  1019--1021 (2003).

\bibitem{Biller:1998hg}
Biller, S.~D., et~al., \emph{Phys. Rev. Lett.}, \textbf{83}, 2108--2111 (1999).

\bibitem{Gleiser:2001rm}
Gleiser, R.~J., and Kozameh, C.~N., \emph{Phys. Rev.}, \textbf{D64}, 083007
  (2001).

\bibitem{HRBL60}
Hughes, V.~W., Robinson, H.~G., and L{\'o}pez, V.~B., \emph{Phys. Rev. Lett.},
  \textbf{4}, 342 (1960).

\bibitem{Drever61}
Drever, R. W.~P., \emph{Phil. Mag.}, \textbf{6}, 683 (1961).

\bibitem{Prestage85}
Prestage, J.~D., Bollinger, J.~J., Itano, W.~M., and Wineland, D.~J.,
  \emph{Phys. Rev. Lett.}, \textbf{54}, 2387 (1985).

\bibitem{Lamoreaux89}
Lamoreaux, S.~K., Jacobs, J.~P., Heckel, B.~R., Raab, F.~J., and Fortson,
  E.~N., \emph{Phys. Rev. D}, \textbf{39}, 1082 (1989).

\bibitem{Chupp89}
Chupp, T.~E., Hoare, R.~J., Loveman, R.~A., Oteiza, E.~R., Richardson, J.~M.,
  Wangshul, M.~E., and Thompson, A.~K., \emph{Phys. Rev. Lett.}, \textbf{63},
  1541 (1989).

\bibitem{Phillips01:LIV_H}
Phillips, D.~F., Humphrey, M.~A., Mattison, E.~M., Stoner, R.~E., Vessot, R.
  F.~C., and Walsworth, R.~L., \emph{Phys. Rev. D}, \textbf{63}, 111101 (2001).

\bibitem{Berglund95}
Berglund, C.~J., Hunter, L.~R., D.~Krause, J., Prigge, E.~O., Ronfeldt, M.~S.,
  and Lamoreaux, S.~K., \emph{Phys. Rev. Lett.}, \textbf{75}, 1879 (1995).

\bibitem{Bear:2000cd}
Bear, D., Stoner, R.~E., Walsworth, R.~L., Kostelecky, V.~A., and Lane, C.~D.,
  \emph{Phys. Rev. Lett.}, \textbf{85}, 5038--5041 (2000).

\bibitem{Cane:2003wp}
Cane, F., et~al., \emph{Phys. Rev. Lett.}, \textbf{93}, 230801 (2004).

\bibitem{Greisen:1966jv}
Greisen, K., \emph{Phys. Rev. Lett.}, \textbf{16}, 748--750 (1966).

\bibitem{Zatsepin:1966jv}
Zatsepin, G.~T., and Kuzmin, V.~A., \emph{JETP Lett.}, \textbf{4}, 78--80
  (1966).

\bibitem{Scully:2000dr}
Scully, S.~T., and Stecker, F.~W., \emph{Astropart. Phys.}, \textbf{16},
  271--276 (2002).

\bibitem{Stecker:2002fh}
Stecker, F.~W., Astrophysics at the highest energy frontiers (2002),
  \texttt{astro-ph/0208507}.

\bibitem{Takeda:1998ps}
Takeda, M., et~al., \emph{Phys. Rev. Lett.}, \textbf{81}, 1163--1166 (1998).

\bibitem{Takeda:1999sg}
Takeda, M., et~al., \emph{Astrophys. J.}, \textbf{522}, 225--237 (1999).

\bibitem{Alfaro:2001gk}
Alfaro, J., and Palma, G., \emph{Phys. Rev.}, \textbf{D65}, 103516 (2002).

\bibitem{Dehmelt:1999jh}
Dehmelt, H., Mittleman, R., van Dyck, J., R.~S., and Schwinberg, P.,
  \emph{Phys. Rev. Lett.}, \textbf{83}, 4694--4696 (1999).

\bibitem{Mittleman:1999it}
Mittleman, R.~K., Ioannou, I.~I., Dehmelt, H.~G., and Russell, N., \emph{Phys.
  Rev. Lett.}, \textbf{83}, 2116--2119 (1999).

\bibitem{Gabrielse:1999kc}
Gabrielse, G., et~al., \emph{Phys. Rev. Lett.}, \textbf{82}, 3198--3201 (1999).

\bibitem{Heckel:1999sy}
Heckel, B.~R., Adelberger, E.~G., Gundlach, J.~H., Harris, M.~G., and Swanson,
  H.~E., Torsion balance test of spin coupled forces, prepared for
  International Conference on Orbis Scientiae 1999: Quantum Gravity,
  Generalized Theory of Gravitation and Superstring Theory Based Unification
  (28th Conference on High-Energy Physics and Cosmology Since 1964), Coral
  Gables, Florida, 16-19 Dec 1999.

\bibitem{Carroll:2001ws}
Carroll, S.~M., Harvey, J.~A., Kostelecky, V.~A., Lane, C.~D., and Okamoto, T.,
  \emph{Phys. Rev. Lett.}, \textbf{87}, 141601 (2001).

\bibitem{Burgess:2002tb}
Burgess, C.~P., Cline, J., Filotas, E., Matias, J., and Moore, G.~D.,
  \emph{JHEP}, \textbf{03}, 043 (2002).

\bibitem{Anisimov:2001zc}
Anisimov, A., Banks, T., Dine, M., and Graesser, M., \emph{Phys. Rev.},
  \textbf{D65}, 085032 (2002).

\bibitem{Weinberg73}
Weinberg, S., \emph{Phys. Rev. D}, \textbf{8}, 4482 (1973).

\bibitem{Susskind:1978ms}
Susskind, L., \emph{Phys. Rev.}, \textbf{D20}, 2619--2625 (1979).

\bibitem{Weinberg:1988cp}
Weinberg, S., \emph{Rev. Mod. Phys.}, \textbf{61}, 1--23 (1989).

\bibitem{Kostelecky:1994rn}
Kostelecky, V.~A., and Potting, R., \emph{Phys. Rev.}, \textbf{D51}, 3923--3935
  (1995).

\bibitem{Coleman:1967ad}
Coleman, S.~R., and Mandula, J., \emph{Phys. Rev.}, \textbf{159}, 1251--1256
  (1967).

\bibitem{Dowker:2003hb}
Dowker, F., Henson, J., and Sorkin, R.~D., \emph{Mod. Phys. Lett.},
  \textbf{A19}, 1829--1840 (2004).

\bibitem{Rovelli:2002vp}
Rovelli, C., and Speziale, S., \emph{Phys. Rev.}, \textbf{D67}, 064019 (2003).

\end{thebibliography}





\end{document}